\documentclass[graybox]{svmult}

\usepackage{mathptmx}       
\usepackage{helvet}         
\usepackage{courier}        
\usepackage{type1cm}        
\usepackage{makeidx}         
\usepackage{graphicx}        
\usepackage{multicol}        
\usepackage[bottom]{footmisc}
\usepackage{hyperref}

\newcommand{\Rpal}{\mathcal{R}}
\newcommand{\LL}{\mathcal{L}}

\makeindex             

\begin{document}

\title*{Nonsingular black holes in Palatini extensions of General Relativity}
\author{Gonzalo J. Olmo}
\institute{Gonzalo J. Olmo \at Departamento de F\'{i}sica Te\'{o}rica and IFIC, Centro Mixto Universidad de Valencia - CSIC. \\ 
Universidad de Valencia, Burjassot-46100, Valencia (Spain), and   \\ 
 Departamento de F\'isica, Universidade Federal da
Para\'\i ba, 58051-900 Jo\~ao Pessoa, Para\'\i ba, Brazil \email{gonzalo.olmo@uv.es}}
%
%
\maketitle

\abstract{An introduction to extended theories of gravity formulated in metric-affine (or Palatini) spaces is presented. Focusing on spherically symmetric configurations with electric fields, we will see that in these theories the central singularity  present in General Relativity is generically replaced by a wormhole structure. The resulting space-time becomes geodesically complete and, therefore, can be regarded as non-singular. We illustrate these properties considering two different models, namely, a quadratic $f(R)$ theory and a Born-Infeld like gravity theory. }

\section{Introduction}
\label{sec:Intro}

Shortly after the publication of Einstein's equations for the gravitational field, Karl Schwarzschild found an exact solution describing the vacuum region surrounding a spherical body of mass $M$. The line element characterizing this space-time takes the form
\begin{equation}
ds^2=-\left(1-\frac{2M}{r}\right)dt^2+\frac{1}{\left(1-\frac{2M}{r}\right)}dr^2+r^2d\Omega^2
\end{equation}
where $r$ is the radial coordinate and  $d\Omega^2\equiv d\theta^2+\sin\theta^2d\varphi^2$ represents the spherical sector. Given the smallness of the quantity $r_S\equiv 2M$, which for a star like the sun is about $r_S\sim 3$km, and the limited astrophysical knowledge about compact objects at that time, this line element was thought to be physically meaningful only in the exterior regions of stars. With the discovery of neutron stars, the physical existence of ultra compact objects was reconsidered and in the 1960's it was understood that geometries such as Schwarzschild's could be a physical reality. In fact, using powerful mathematical techniques it was concluded that under reasonable conditions, complete gravitational collapse is unavoidable for sufficiently massive objects \cite{Theo1,Theo2,Theo3,Theo4,Theo5}. Black holes, therefore, are an important prediction of Einstein's theory of General Relativity (GR). \\

The existence of black holes has a deep impact for the theoretical consistency of GR. In fact, given that the laws of Physics as we know them are defined on top of a dynamical geometry, the space-time, if the geometry becomes ill defined at some event then our ability to describe physical phenomena and make predictions will be seriously affected \cite{Hawking76}. This is precisely what happens in the interior of black holes. 

In the Schwarzschild case, for instance, any observer within the region $r<r_S$ is forced to travel towards decreasing values of $r$, being $r=0$ reached in a finite proper time \cite{MTW}. At that location, curvature scalars diverge and gravitational forces are so strong that any extended body is instantaneously crushed to zero volume. Thus, any observer reaching $r=0$ is destroyed and disappears together with its ability to describe the physical processes taking place in that region. Under this circumstance, it is typically stated that the Schwarzschild black hole contains a singularity or that it describes a singular space-time. \\

The notion of singularity is a very elusive concept, though \cite{Curiel2009}. The Schwarzschild example suggests that curvature divergences can somehow be regarded as a signature of their existence. However, if one takes a space-time such as Minkowski and artificially removes a portion of it, any observer or signal that propagates through it and reaches the boundary of the removed portion simply vanishes there, as there is nowhere to go {\it beyond} that boundary. One can also find observer trajectories which intersect this boundary in their past, suggesting that they came into existence out of the blue. The potential creation and/or destruction of physical observers and/or light signals in a given space-time is thus fundamental to determine if an appropriate physical description is possible or not. For this reason, for the characterization of singular space-times one should not  focus on the potential existence of infinities in the gravitational fields, which are absent in the amputated Minkowskian example, but rather one should be worried about the existence of physical observers at all times. 

Following this line of reasoning, it is generally stated that a singular space-time is one in which there exist incomplete timelike and/or null geodesics, i.e., geodesics which cannot be extended to arbitrary values of their affine parameter in the past or in the future \cite{Geroch:1968ut,Hawking:1973uf,Wald:1984rg} (see also \cite{Senovilla:2014gza} for a more recent discussion of this point and references). Note, in this sense, that observers are identified with geodesic curves. The incompleteness of geodesics, therefore, hinges in the fact that in order to be able to provide a reliable description of phenomena on a given space-time, physical observers and/or signals should never be created or destroyed, i.e., their existence should be unrestricted along their worldline. The presence of curvature divergences is thus irrelevant for the determination of whether a space-time is singular or not: the potential {\it suffering} of observers due to intense tidal forces is not comparable to the importance of their very existence. 

The fact that the Schwarzschild solution, as well as all other black hole solutions known to date, represent  geodesically incomplete space-times is thus a serious conceptual limitation of GR. Improvements in the theory are thus necessary, which has motivated different approaches to the problem of singularities. Some of those  are based on the idea of bounded curvature scalars \cite{Mukhanov:1991zn, Ansoldi:2008jw,Lemos, Spallucci, Bronnikov,Hayward} which, however, is logically unrelated to the notion of geodesic completeness. \\

  In these lectures we will be dealing with certain (classical) extensions of GR in which simple non-rotating black hole solutions which are geodesically complete, and hence nonsingular, are possible. The approach presented here does not follow the intuitive and widespread idea that to get a nonsingular theory one should keep curvature scalars bounded. In our case, curvature divergences do arise in some regions but their presence is not an obstacle to have complete geodesic paths\footnote{This provides a counterexample to the correlation typically observed in GR between space-times with incomplete geodesics and which contain curvature divergences.} \cite{Olmo:2015bya}. Making a long story short, this is accomplished by the replacement of the black hole center by a wormhole \cite{wormhole,Lobo:2007zb}. Unlike the case of GR, in our approach one does not need exotic matter sources to generate the wormhole.  Rather, a simple free electric field will be able to do the job. Also, our geometries are not designed {\it a priori} but, rather, follow directly by integrating the field equations once the matter fields are specified. It is in this sense that these wormholes are more natural than those typically discussed in the context of GR, where one first defines the metric and then obtains the necessary stress-energy tensor by plugging it in Einstein's equations. \\

It is worth noting at this point that the use of nontrivial topologies (wormholes) in combination with self-gravitating free fields as a way to cure space-time singularities was suggested long ago by J.A. Wheeler \cite{Wheeler:1955zz}. We will see that our solutions represent an explicit example of geons in Wheeler's sense \cite{Lobo:2013adx,Lobo:2013vga} and, as such, avoid the well-known {\it problem of the sources} \cite{Ortin} that one finds in GR for the Schwarzschild and Reissner-Nordstr\"{o}m black holes, for instance. \\

The content is organized as follows. In Sec. \ref{sec:framework} our geometrical scenario is introduced, making emphasis on the importance of understanding gravitation as a geometric phenomenon and geometry as an issue of metrics and connections, i.e., as something else than a theory of just metrics. Once the fundamental notions of metric-affine geometry have been presented, in Sec. \ref{sec:GR} we work out the field equations of GR \`{a} la Palatini, and in Sec.
\ref{sec:EOM}  we do the same for two models of interest, namely, a quadratic $f(R)$ theory and a Born-Infeld-like gravity theory. The first example appears naturally in that quadratic corrections in curvature are common to many different approaches to quantum and non-quantum extensions of GR. The simplicity of this model comes at the price of introducing a nonlinear theory of electrodynamics as matter source in order to obtain the desired effects in the equations. The Born-Infeld case, on the contrary, can be easily combined with a standard Maxwell electric field. In both cases, exact analytical black hole solutions can be found, which allows us to explore the behavior of geodesics in both geometries in detail. 
The equations governing black hole structure are derived in a generic form in Sec.  \ref{sec:generic} and applied to the gravitational Born-Infeld model in Sec.\ref{sec:BI} and to the $f(R)$ model in Sec. \ref{sec:f(R)}. The study of geodesics appears in Sec. \ref{sec:geodesics}. We conclude in Sec. \ref{sec:theend} with a brief summary and discussion of the results.

\section{Basic framework: metric-affine gravity. }\label{sec:framework}

In elementary courses on gravitation \cite{MTW} one learns that general covariance is accomplished by replacing flat Minkowskian derivatives $\partial_\mu$ by covariant derivatives $\nabla_\mu$, whose action on vector components (for instance) is of the form $\nabla_\mu A_\nu=\partial_\mu A_\nu-\Gamma^\lambda_{\mu\nu}A_\lambda$. Here  $\Gamma^\lambda_{\mu\nu}$ is the so-called Levi-Civita connection, which is defined as 
\begin{equation}\label{eq:LC}
\Gamma^\lambda_{\mu\nu}=\frac{g^{\lambda\rho}}{2}\left[\partial_\mu g_{\rho\nu}+\partial_\nu g_{\rho\mu}-\partial_\rho g_{\mu\nu}\right] \ ,
\end{equation}
with $g_{\mu\nu}$ representing the space-time metric. 
The connection has a non-tensorial transformation law which compensates the action of $\partial_\mu$ in such a way that $\nabla_\mu A_\nu$ transforms as a tensor under arbitrary changes of coordinates. With the connection one defines the Riemann curvature tensor as 
\begin{equation}\label{eq:Rie}
{R^\alpha}_{\beta\mu\nu}= \partial_\mu \Gamma^\alpha_{\nu\beta}-\partial_\nu \Gamma^\alpha_{\mu\beta}+ \Gamma^\kappa_{\nu\beta} \Gamma^\alpha_{\mu\kappa}- \Gamma^\kappa_{\mu\beta} \Gamma^\alpha_{\nu\kappa} \ ,
\end{equation}
and Einstein's equations take the form 
\begin{equation}\label{eq:GR}
R_{\beta\nu}-\frac{1}{2}g_{\beta\nu}R=\kappa^2 T_{\beta\nu} \ , 
\end{equation}
where $R_{\beta\nu}={R^\lambda}_{\beta\lambda\nu}$ is the Ricci tensor, $R=g^{\mu\nu} R_{\mu\nu}$ the Ricci curvature scalar,  $T_{\beta\nu}$ the stress-energy tensor of the matter, and $\kappa^2=8\pi G/c^4$. Written in this form, GR is a theory based on the metric tensor $g_{\mu\nu}$ as the field that describes gravitational interactions.\\

Interestingly, at the time Einstein formulated GR, the theory of affine connections had not been developed yet. Only Riemannian geometry, based on the metric tensor, was available to implement his idea of gravitation as a geometric phenomenon. Einstein's theory boosted the interest of mathematicians on differential geometry, giving rise to the study of non-Riemannian spaces \cite{Eisenhart}. It was then established that general covariance could be implemented without defining a metric structure. This is so because the non-tensorial transformation law of the connection is a property that does not depend on the particular form of the connection, i.e., it is independent of the definition (\ref{eq:LC}). As a consequence, the Riemann curvature tensor (\ref{eq:Rie}) can be defined without referring it to a metric. 

This point is very important because it opens a whole new range of possibilities to implement the idea of gravitation as a geometric phenomenon. Is the space-time geometry Riemannian? It is rather apparent that the Euclidean space of Newtonian mechanics is not appropriate to describe relativistic phenomena, but that does not lead uniquely to the Riemannian case ({\it the metric as the foundation of all}). Whether the space-time geometry is Riemannian or not is a fundamental question that must be answered by experiments, as Einstein himself stated \cite{Feigl}. We must, obviously, admit that the Riemannian description of GR is very successful at the length scales and energies accessible in laboratory and the Solar system (as well as in other systems whose orbital motions are well understood) \cite{Will:2014kxa}. However, there is still a broad range of energies and length scales that lie beyond direct experimental scrutiny. Demanding that the Riemannian condition (\ref{eq:LC}) , or $\nabla_\mu g_{\alpha\beta}=0$,
 be satisfied at all scales might be an excessive assumption/constraint. \\

Aside from the purely theoretical interest in non-Riemannian geometries, there are other reasons to explore the effects that independent metric and affine degrees of freedom could have in gravitation. It turns out that in continuous systems with an ordered microstructure, such as in Bravais crystals or materials as popular as graphene, one needs a metric-affine geometry in order to correctly describe macroscopic properties like viscosity or plasticity \cite{Kroner1,Kleinert}. These properties are intimately related with the existence of defects in the microstructure. And these defects are responsible for the independence between metric and affine degrees of freedom. For instance, in a crystal without defects, one can introduce a notion of distance (metricity) by counting atoms along crystalographic directions (a special set of directions in the structure which minimize distances) \cite{Kroner1,Kroner2,RdWitt,Kroner3,Kroner4}. However, if there exist point defects such as missing atoms, the microscopic process of step counting breaks down and the idea of metricity cannot be translated to the continuum in any natural way. 

The microscopic notion of distance can be extended to the continuum by defining an {\it auxiliary or idealized} structure without defects in which the step-counting procedure is naturally implemented. Physical distances can be defined once the density of defects is known, which allows to establish a correspondence between the idealized structure and the physical one. The idealized crystalographic directions need not coincide everywhere with the directions that minimize physical distances, which implies that the physical metric $g_{\alpha\beta}$ is not conserved along the idealized paths, i.e.,   $\nabla_\mu^{(\Gamma)} g_{\alpha\beta}\neq 0$, where $\Gamma$ is the connection associated to the auxiliary metric. The quantity $Q_{\mu\alpha\beta}\equiv \nabla_\mu^{(\Gamma)} g_{\alpha\beta}$, known as non-metricity tensor, then plays a relevant role in the physical description of the continuized system. 

Another interesting geometric structure arises when there exist dislocations (one-dimensional defects). It is well-known that dislocations are the discrete version of torsion \cite{Kondo, Bilby}. Crystals with a certain density of dislocations, therefore, lead to effective geometries with a metric and a non-symmetric connection, which is related to the Einstein-Cartan theory of gravity \cite{Kleinert}. Given that point defects (vacancies and intersticials) can interact with dislocations (creating and/or destroying them), a complete theory should have into account the metric, the non-metricity tensor, and the torsion. If the space-time had a microstructure with defects, such as that suggested by the notion of space-time foam, the continuum that we perceive could require geometric structures beyond those typically considered in Einstein's theory of gravity \cite{Lobo:2014nwa,Olmo:2015bha, Olmo:2015wwa}. \\

It is for the above simple reasons that we are going to explore several examples of theories of gravity assuming that metric and connection are equally fundamental and {\it a priori} independent fields. Imposing a {\it principle of democracy}, we will derive the equations governing the metric and the connection from an action, without imposing any {\it a priori} constraint between them. The field equations should determine how metric and affine degrees of freedom interact between them and with the matter fields.

\section{General Relativity \`{a} la Palatini}\label{sec:GR}
  
To begin with, it is useful to consider the metric-affine or Palatini version of GR \cite{Origin}. The action functional for the Einstein-Palatini theory can be written as 
\begin{equation}
S=\frac{1}{2\kappa^2}\int d^4x \sqrt{-g} g^{\mu\nu}R_{\mu\nu}(\Gamma)+S_m(g_{\mu\nu},\psi) \ ,
\end{equation}
where $R_{\mu\nu}(\Gamma)={R^\alpha}_{\mu\alpha\nu}$ is defined in terms of a connection which is {\it a priori} independent of the metric $g_{\mu\nu}$, $S_m$ represents the matter action, and $\psi$ denotes collectively the matter fields\footnote{For simplicity, in the matter action we have only assumed a dependence on the metric. This prescription is compatible with the experimental evidence on the Einstein equivalence principle \cite{Will:2014kxa}. However, dependence on the connection should also be allowed to explore its phenomenology in regimes not yet accessed experimentally. The coupling of fermions to gravity, whose spin may source the torsion tensor (antisymmetric part of the connection), is a particular case of interest which has been considered explicitly in supergravity theories and in the Einstein-Cartan theory \cite{Ortin}, for example.}. 

Variation of the action with respect to the (inverse) metric and the connection leads to 
\begin{equation}\label{eq:varGR0}
\delta S=\frac{1}{2\kappa^2}\int d^4x \sqrt{-g}\left[\left(R_{\mu\nu}(\Gamma)-\frac{1}{2} g_{\mu\nu}g^{\alpha\beta}R_{\alpha\beta}(\Gamma)-\kappa^2T_{\mu\nu}\right)\delta g^{\mu\nu}+g^{\mu\nu}\delta R_{\mu\nu}\right] \ ,
\end{equation}
where 
\begin{equation}
\delta R_{\mu\nu}= \nabla_\lambda\left(\delta \Gamma^\lambda_{\nu\mu}\right)-\nabla_\nu\left(\delta \Gamma^\lambda_{\lambda\mu}\right)+2S^\rho_{\alpha\nu}\delta \Gamma^\alpha_{\rho\mu} \ ,
\end{equation}
and $S^\rho_{\alpha\nu}\equiv \frac{1}{2}\left(\Gamma^\rho_{\alpha\nu}-\Gamma^\rho_{\nu\alpha}\right)$ is the torsion tensor. For simplicity, in the following derivations we will skip all torsional terms\footnote{We do this to focus our attention on the symmetric part of the connection but we do admit the possibility of having an antisymmetric part because fermions do exist in Nature. Note in this sense that, in general, assuming a symmetric connection before performing the variations or setting it to zero after the field equations have been obtained are inequivalent procedures. A detailed discussion with concrete examples can be found in  \cite{Olmo:2013lta}.}. After elementary manipulations, and knowing that $\nabla_\mu(\sqrt{-g}J^\mu)=\partial_\mu (\sqrt{-g}J^\mu)+2S^\lambda_{\lambda\mu}(\sqrt{-g}J^\mu)$, Eq.(\ref{eq:varGR0}) turns into
\begin{eqnarray}\label{eq:varGR1}
\delta S&=&\frac{1}{2\kappa^2}\int d^4x \sqrt{-g}\left[\left(R_{\mu\nu}(\Gamma)-\frac{1}{2} g_{\mu\nu}g^{\alpha\beta}R_{\alpha\beta}(\Gamma)-\kappa^2T_{\mu\nu}\right)\delta g^{\mu\nu}\right.\nonumber \\ 
&+& \left.\left(-\nabla_\lambda\left(\sqrt{-g}g^{\mu\nu}\right)+\delta^\mu_\lambda\nabla_\rho\left(\sqrt{-g}g^{\rho\nu}\right)\right)\delta \Gamma^\lambda_{\mu\nu}\right] \ .
\end{eqnarray}
The field equations are obtained by setting to zero the coefficients multiplying the independent variations $\delta g^{\mu\nu}$ and $\delta\Gamma^\lambda_{\mu\nu}$, which yields
\begin{eqnarray}
R_{\mu\nu}(\Gamma)-\frac{1}{2} g_{\mu\nu}g^{\alpha\beta}R_{\alpha\beta}(\Gamma)&=& \kappa^2T_{\mu\nu}\label{eq:metvarGR}\\
-\nabla_\lambda\left(\sqrt{-g}g^{\mu\nu}\right)+\delta^\mu_\lambda\nabla_\rho\left(\sqrt{-g}g^{\rho\nu}\right)&=&0 \ . \label{eq:convarGR}
\end{eqnarray}
Contracting the indices $\mu$ and $\lambda$ in (\ref{eq:convarGR}) one finds that 
$\nabla_\rho\left(\sqrt{-g}g^{\rho\nu}\right)=0$, which turns that equation into
\begin{equation}\label{eq:convarGR1a}
\nabla_\lambda\left(\sqrt{-g}g^{\mu\nu}\right)=0 \ .
\end{equation}
Writting this equation explicitly, we get
\begin{equation}\label{eq:convarGR1b}
g^{\mu\nu}\partial_\lambda \sqrt{-g}+\sqrt{-g}\partial_\lambda g^{\mu\nu}+\sqrt{-g}\left[-\Gamma^\alpha_{\alpha\lambda}g^{\mu\nu}+\Gamma^\mu_{\lambda\alpha}g^{\alpha\nu}+\Gamma^\nu_{\lambda\alpha}g^{\alpha\mu}\right]=0 \ ,
\end{equation}
and contracting with $g_{\mu\nu}$ we find that $\Gamma^\alpha_{\alpha\mu}=\partial_\mu \ln \sqrt{-g}$, where the relation $g_{\mu\nu}\partial_\lambda g^{\mu\nu}=-2\partial_\lambda \ln \sqrt{-g}$ has been used. Inserting this result in (\ref{eq:convarGR1b}), one finds that (\ref{eq:convarGR1a}) is equivalent to $\nabla_\lambda g^{\mu\nu}=0$. Given that $g_{\mu\rho}g^{\rho\nu}=\delta_\mu^\nu$, one readily verifies that  $\nabla_\lambda g^{\mu\nu}=0$ also implies $\nabla_\lambda g_{\mu\nu}=0$. This last relation can be used to obtain the form of $\Gamma^\alpha_{\mu\nu}$ as a function of the metric and its first derivatives by just using algebraic manipulations \cite{Olmo:2012yv}. The result is simply that $\Gamma^\alpha_{\mu\nu}$ boils down to the Levi-Civita connection defined in (\ref{eq:LC}). As a consequence, the Ricci tensor $R_{\mu\nu}(\Gamma)$ turns into the Ricci tensor of the metric $g_{\mu\nu}$ and (\ref{eq:metvarGR}) coincides with the Einstein equations (\ref{eq:GR}). 

In summary, the Einstein-Palatini action exactly recovers Einstein's equations (in the torsionless case) and implies that the geometry is Riemannian without the need of imposing the {\it compatibility condition} $\nabla_\lambda g_{\mu\nu}=0$ as an input. 

It is important to remark at this point that the constraint  $\nabla_\lambda g_{\mu\nu}=0$ between metric and connection is a property that belongs naturally to the Einstein-Palatini theory but which is not {\it a priori} guaranteed in other theories. Nonetheless, in most of the literature on extended theories of gravity it has been implicitly assumed as true, forcing the geometry to be Riemannian from the onset (see, however, \cite{Olmo:2011uz} for a review on Palatini gravity). We will see in the following that relaxing this constraint and allowing the theory to determine the form of the connection from a variational principle, the compatibility between metric and connection is generically lost. The implications of this will be nontrivial, providing new phenomenology that will be relevant in the study of black hole interiors. 

\section{Beyond GR} \label{sec:EOM}

Considering extensions of GR to address questions concerning high and very high energies one naturally finds the possibility of adding quadratic and/or higher order curvature corrections in the gravitational Lagrangian. Such corrections arise when one considers quantum fields propagating in curved space-times \cite{QFTCS1,QFTCS2}, in the low-energy limits of string theories \cite{Ortin}, and in effective field theory or  phenomenological approaches \cite{Cembranos:2008gj,Schimming:2004yx}. Theories such as  $R+\lambda R^2+\gamma R_{\mu\nu} R^{\mu\nu} + \beta {R^\alpha}_{\beta\mu\nu}{R_\alpha}^{\beta\mu\nu}$, for instance, have been typically  considered in the literature on the early universe and in black hole scenarios \cite{Q1,Q2,Q3,Q4,Q5,Q6,Starobinsky,Anderson,Economou:1993va,LuPope,Gullu:2011sj,Liu:2011kf}. The Riemann-squared dependence is typically removed because it can be combined with the other quadratic terms to give the so-called Gauss-Bonnet term, which does not contribute to the field equations and simply redefines the coefficients $\lambda$ and $\gamma$. 

The standard argument is that high-order curvature corrections could capture some relevant new physics beyond the range of applicability of GR but below the full quantum gravity regime.  Given the higher-order character of the resulting field equations, analytical solutions are hard to find in general. Numerical solutions do exist and regular cases (in the sense of bounded curvature scalars \cite{Ansoldi:2008jw}) have been found for static black hole configurations \cite{Berej:2006cc} coupled to nonlinear theories of electrodynamics using perturbative methods. \\

The extensive literature existing on the metric (or Riemannian) formulation of quadratic gravity contrasts with the little attention received by its metric-affine counterpart. Interestingly, through recent work carried out in the last years, it has been established that in the Palatini version of those theories one always finds analytical solutions \cite{or12a,or12b,or12c}. In the following we will study the field equations of models similar to the quadratic theory  mentioned above but formulated in the Palatini approach. We will then focus on spherically symmetric configurations in which new black hole solutions can be found. 
 
\subsection{$f(R)$ theories}\label{sec:f(R)}

The derivation of the field equations for theories of the $f(\Rpal)$ type, where $f$ represents a certain function of the Ricci scalar\footnote{The typography $\Rpal$ is used here to emphasize that this scalar is built by combining the metric $g_{\mu\nu}$ with the Ricci tensor of a connection $\Gamma^\alpha_{\mu\nu}$ whose relation with $g_{\mu\nu}$ is {\it a priori} unknown. Whenever $\Gamma^\alpha_{\mu\nu}$ be defined in terms of a metric $k_{\mu\nu}$, then we will use the notation $R(k)=k^{\mu\nu}R_{\mu\nu}(k)$.} $\Rpal=g^{\mu\nu}R_{\mu\nu}(\Gamma)$, is straightforward and follows essentially the same steps as in the case of GR presented in Sec. \ref{sec:GR}. Variation of the action leads to the equations (see, for instance, \cite{Olmo:2012yv,Olmo:2011uz} for details) 
\begin{eqnarray}
f_{\Rpal} R_{\mu\nu}(\Gamma)-\frac{1}{2} g_{\mu\nu}f(\Rpal)&=& \kappa^2T_{\mu\nu}\label{eq:metvarfR}\\
-\nabla_\lambda\left(\sqrt{-g}f_{\Rpal} g^{\mu\nu}\right)+\delta^\mu_\lambda\nabla_\rho\left(\sqrt{-g}f_{\Rpal}g^{\rho\nu}\right)&=&0 \ , \label{eq:convarfR}
\end{eqnarray}
where we denote $f_{\Rpal}\equiv df/d{\Rpal}$. Manipulating the connection equation (\ref{eq:convarfR}), one finds that it can be reduced to 
\begin{equation}\label{eq:convarfR1}
\nabla_\lambda\left(\sqrt{-g}f_{\Rpal} g^{\mu\nu}\right)=0 \ .
\end{equation}
Before proceeding with further manipulations, it is important to interpret this equation in combination with (\ref{eq:metvarfR}). At first sight, one may think that (\ref{eq:convarfR1}) contains up to second order derivatives of the connection because $f_{\Rpal}$ is being acted upon by a derivative operator and it already contains first-order derivatives of $\Gamma^\alpha_{\mu\nu}$ via its dependence on $\Rpal$. However, taking the trace of (\ref{eq:metvarfR}) with $g^{\mu\nu}$, one finds the important relation
\begin{equation}\label{eq:tracefR}
\Rpal f_{\Rpal}-2f=\kappa^2 T \ ,
\end{equation}
which establishes an algebraic relation between $\Rpal$ and $T$, generalizing in this way the case $\Rpal=-\kappa^2 T$ to nonlinear Lagrangians. This allows us to reinterpret (\ref{eq:convarfR1}) as an equation in which the independent connection $\Gamma^\alpha_{\mu\nu}$  satisfies an algebraic linear equation which involves the matter fields through the function $f_{\Rpal}$ and the metric. 

A solution to this equation can be obtained \cite{Olmo:2009xy} by considering the existence of a rank-two tensor $h_{\mu\nu}$ such that $ \sqrt{-g}f_{\Rpal} g^{\mu\nu}$ can be written as $\sqrt{-h}h^{\mu\nu}$. With this identification, Eq. (\ref{eq:convarfR1}) turns into $\nabla_\mu (\sqrt{-h}h^{\alpha\beta})=0$, with $h_{\mu\nu}=f_{\Rpal} g_{\mu\nu}$, and the solution can be obtained in much the same way as in the GR case (see the manipulations following Eq.(\ref{eq:convarGR1a})). As a result, we find that $\Gamma^\alpha_{\mu\nu}$ can be written as the Levi-Civita connection of the {\it auxiliary metric} $h_{\mu\nu}$, i.e., 
\begin{equation}\label{eq:LCfR}
\Gamma^\lambda_{\mu\nu}=\frac{h^{\lambda\rho}}{2}\left[\partial_\mu h_{\rho\nu}+\partial_\nu h_{\rho\mu}-\partial_\rho h_{\mu\nu}\right] \ .
\end{equation}
This result is valid for any Palatini theory of the $f(\Rpal)$ type, including GR. \\

We now turn our attention to the metric field equations (\ref{eq:metvarfR}), which contains elements referred to the metric $g_{\mu\nu}$ and others, like $R_{\mu\nu}(\Gamma)$, that depend on $h_{\mu\nu}$. Given that $g_{\mu\nu}=(1/f_{\Rpal})h_{\mu\nu}$ are conformally related, one can express $R_{\mu\nu}(\Gamma)$ in terms of $R_{\mu\nu}(g)$ and derivatives of $f_{\Rpal}$ using well-known formulas \cite{Olmo:2005hc,Olmo:2005zr} (see, for instance, Appendix D in Wald's book \cite{Wald:1984rg}). Another possibility is to express everything in terms of $h_{\mu\nu}$. This is the approach we will follow because it leads to a very compact expression of the form
\begin{equation}\label{eq:FE-fR}
{R^\mu}_\nu(h)=\frac{\kappa^2}{f_{\Rpal}^2}\left[\frac{f}{2\kappa^2}{\delta^\mu}_\nu+{T^\mu}_\nu\right] \ ,
\end{equation}
where ${R^\mu}_\nu(h)=h^{\mu\lambda}R_{\lambda\nu}(h)$ and ${T^\mu}_\nu=g^{\mu\lambda}T_{\lambda\nu}$. Written in this form, it is apparent that the auxiliary metric $h_{\mu\nu}$ satisfies a set of second-order equations with a structure very similar to that found in GR. In fact, on the left-hand side we find a second-order differential operator acting on $h_{\mu\nu}$, whereas on the right-hand side we have the matter, represented by ${T_\mu}^\nu$ and by $f$ and $f_{\Rpal}$, which are both functions of the trace $T$ of ${T_\mu}^\nu$. 

With the equations written in this form, one may try to solve for $h_{\mu\nu}$ and then obtain $g_{\mu\nu}$ by just using the conformal relation $g_{\mu\nu}=(1/f_{\Rpal})h_{\mu\nu}$. This strategy might not always be straightforward, but will be very useful in the cases we will be dealing with. \\

To conclude with the discussion of $f(R)$ theories, it is important to consider the vacuum solutions. Such solutions correspond to the case in which ${T_\mu}^\nu=0$, which implies $T=0$. As a result, the algebraic equation (\ref{eq:tracefR}) implies $\Rpal=\Rpal_{vac}$, where $\Rpal_{vac}$ is some constant which may depend on the parameters that characterize the specific $f(\Rpal)$ Lagrangian chosen (obviously, some models may yield more than one solution and the good ones should be selected on physically reasonable grounds). A constant $\Rpal$ implies that any function of $\Rpal$ is also a constant. A direct consequence of this is that the conformal factor relating $g_{\mu\nu}$ and $h_{\mu\nu}$ can be absorbed into an irrelevant redefinition of units, making the two metrics coincide. This means that in vacuum the connection (\ref{eq:LCfR}) boils down to the Levi-Civita connection of $g_{\mu\nu}$. Also, the metric field equations (\ref{eq:FE-fR}) recover the equations of GR in vacuum, with an effective cosmological constant. All this implies that the vacuum solutions of the theory are exactly the same as those appearing in vacuum GR (although different boundary conditions may apply).  Therefore, in order to explore new physics beyond GR, one must consider explicitly the presence of matter sources. In this sense, we note that though the Schwarzschild solution is a mathematically acceptable solution of all Palatini $f(R)$ theories in vacuum, one should carefully consider the boundary conditions necessary to match that solution with the solution in the region containing the sources. The intuitive view that a delta-like distribution at the center is valid is not guaranteed here, as some models exhibit upper bounds for the density and pressure \cite{Olmo:2009xy,MartinezAsencio:2012xn}. For this reason, vacuum solutions must be handled with care, and non-vacuum solutions should be explored to gain insight on the properties of these theories. 

\subsection{Born-Infeld gravity}

The Born-Infeld gravity model is defined by means of the following action
\begin{equation}\label{eq:BI0}
S=\frac{1}{\kappa^2\epsilon}\int d^4x \left[\sqrt{-|g_{\mu\nu}+\epsilon R_{\mu\nu}|}-\lambda \sqrt{-|g_{\mu\nu}|}\right]+S_m[g_{\mu\nu},\psi] \ ,
\end{equation}
where  vertical bars inside the square-root denote the determinant of that quantity, and $\epsilon$ is a small parameter with dimensions of length squared. This model was first consider in metric formalism \cite{Deser:1998rj}, where the model suffers from a ghost instability due to its nonlinear dependence on the Ricci tensor. In \cite{Vollick:2003qp}, the theory was studied within the Palatini formalism, finding that in that approach the ghost is avoided. The phenomenological consequences of this theory have since then been extensively explored  in cosmology \cite{Du:2014jka, Kim:2013noa,Kruglov:2013qaa, Yang:2013hsa,Avelino:2012ue, DeFelice:2012hq, EscamillaRivera:2012vz, Cho:2012vg, Scargill:2012kg, EscamillaRivera:2013hv,Banados}, astrophysics \cite{Harko:2013xma,Avelino:2012ge}, stellar structure \cite{Sham:2013cya,Kim:2013nna,Harko:2013wka,Sham:2013sya, Avelino:2012qe, Sham:2012qi,Pani:2012qd,  Pani:2011mg}, the problem of cosmic singularities \cite{Bouhmadi-Lopez:2013lha, Ferraro:2010at}, black holes \cite{Olmo:2015dba, Olmo:2013gqa}, and wormhole physics \cite{Shaikh:2015oha,Bambi:2015sla, Lobo:2014fma,Harko:2013aya}, among many others. Extensions of the original formulation have also been considered \cite{Odintsov:2014yaa, Makarenko:2014lxa, Makarenko:2014nca,BI-extensions1,BI-extensions2,BI-extensions3,BI-extensions4,BI-extensions5,BI-extensions6,BI-extensions7,BI-extensions8,BI-extensions9,BI-extensions10, Jimenez:2014fla,Jimenez:2015caa,Jimenez:2015jqa}.  

In the limit $\epsilon\to 0$, this action recovers the quadratic\footnote{As mentioned before, in the quadratic theory the dependence on the Riemann squared term can be eliminated by a simple redefinition of the coefficients in front of $R^2$ and $R_{\mu\nu}R^{\mu\nu}$. It is this Ricci-dependent theory which is recovered from the Born-Infeld action. We also note that the Ricci tensor in the action is symmetric. Though this is not obvious {\it a priori}, it can be shown that it is indeed true when torsion is set to zero at the level of the field equations \cite{Olmo:2013lta}. } gravity theory mentioned at the beginning of this section with specific coefficients in front of $R^2$ and $R_{\mu\nu} R^{\mu\nu}$ 
\cite{Olmo:2013gqa}. The parameter $\lambda$ is related to the cosmological constant, which vanishes if $\lambda=1$. From now on we will set $\lambda=1$ for simplicity. Higher-order contractions of the Ricci tensor arise as higher-order corrections in $\epsilon$ are considered. \\

The derivation of the field equations is straightforward if one introduces the definition 
\begin{equation}\label{eq:BI-h}
h_{\mu\nu}=g_{\mu\nu}+\epsilon R_{\mu\nu} \ ,
\end{equation} 
which allows to express the action (\ref{eq:BI0}) in the more compact form
\begin{equation}\label{eq:BI1}
S=\frac{1}{\kappa^2\epsilon}\int d^4x \left[\sqrt{-h}- \sqrt{-g}\right]+S_m[g_{\mu\nu},\psi] \ .
\end{equation}
Variation of the action with respect to metric and connection\cite{Olmo:2013gqa,Odintsov:2014yaa} leads to
\begin{eqnarray}\label{eq:metvarBI0}
\sqrt{-h}h^{\mu\nu}-\sqrt{-g}g^{\mu\nu}&=&-\epsilon \sqrt{-g}\kappa^2T^{\mu\nu} \\
\nabla_\mu(\sqrt{-h}h^{\alpha\beta})&=&0 \label{eq:convarBI}
\end{eqnarray}
It is clear from (\ref{eq:convarBI}) that one can formally solve for the connection as the Levi-Civita connection of the {\it auxiliary metric} $h_{\mu\nu}$. Accepting that possibility, then we find that on the right-hand side of our original definition (\ref{eq:BI-h}) the Ricci tensor contains up to second-order derivatives of $h_{\mu\nu}$. This simply indicates that to obtain $h_{\mu\nu}$ we need to solve some differential equations which involve $g_{\mu\nu}$ and $R_{\mu\nu}(h)$ . In order to be able to do it, we must first find the relation that exists between $h_{\mu\nu}$ and the pair $(g_{\mu\nu},T_{\mu\nu})$. This relation is determined by Eq. (\ref{eq:metvarBI0}). In fact, assuming that $h_{\mu\nu}$ and $g_{\mu\nu}$ are related by some {\it deformation} matrix in the form
\begin{equation}\label{eq:Omega}
 h_{\mu\nu}=g_{\mu\alpha}{\Omega^\alpha}_\nu  \ , \ h^{\mu\nu}={(\Omega^{-1})^\mu}_\alpha g^{\alpha\nu}  \ ,
\end{equation}
 then we can write (\ref{eq:metvarBI0}) as 
\begin{equation}\label{eq:BI-Omega}
\sqrt{|\Omega|}{(\Omega^{-1})^\mu}_\nu ={\delta^\mu}_\nu-\epsilon \kappa^2 {T^\mu}_\nu \ .
\end{equation} 
This equation tells us that the deformation that relates $h_{\mu\nu}$ with $g_{\mu\nu}$ is determined by the local distribution of energy-momentum. This is similar to what we already observed in the case of $f(R)$ theories, where the conformal factor relating the metrics was a function of the trace of ${T_\mu}^\nu$ [see Eq.(\ref{eq:tracefR})].  Note also that for this model the explicit form of ${\Omega^\alpha}_\nu$ is
\begin{equation}
{\Omega^\alpha}_\nu={\delta^\alpha}_\nu+\epsilon g^{\alpha\beta}R_{\beta \nu}(h) \ .
\end{equation}
Eq.(\ref{eq:BI-Omega}) is thus telling us that the object $g^{\alpha\beta}R_{\beta \nu}(h)$, which is a hybrid tensor that mixes $g^{\alpha\beta}$ with $h_{\mu\nu}$, is an algebraic function of the stress-energy tensor ${T^\mu}_\nu$. This is analogous to the relation between the scalar quantities $\Rpal$ and $T$ in the $f(\Rpal)$ case. \\

Having established the explicit relation between $h_{\mu\nu}$ and $g_{\mu\nu}$, we can now go back to (\ref{eq:BI-h}) and write an equation for $h_{\mu\nu}$ and the matter. With a bit of algebra, one finds that the corresponding equations can be written as
\begin{equation}\label{eq:FE-BI}
{R^\mu}_\nu(h)=\frac{\kappa^2}{\sqrt{|\Omega|}}\left[\frac{\sqrt{|\Omega|}-\lambda}{\kappa^2\epsilon}{\delta^\mu}_\nu+{T^\mu}_\nu\right] \ .
\end{equation}
The structure of these equations is very similar to that found in the case of $f(\Rpal)$ theories, with the Ricci tensor of the metric $h_{\mu\nu}$ on the left-hand side and functions of the matter fields on the right. We will see that in some cases of interest it will be possible to solve for $h_{\mu\nu}$ and then use (\ref{eq:Omega}) to obtain $g_{\mu\nu}$. 

We also note here that the vacuum solutions of this model recover the field equations of vacuum GR. This is clearly seen from Eq.(\ref{eq:BI-h}), which in vacuum implies that the matrix ${\Omega_\mu}^\alpha$ is a constant times the identity (when $\lambda=1$, this constant is just unity). As a result the two metrics are physically equivalent and one recovers the equations of vacuum GR. The exploration of new physics should thus be carried out considering explicitly the presence of matter sources. 

\subsection{Generic field equations}

The field equations obtained in the previous subsections for two different types of gravity models suggests that there exists a basic structure for the field equations in Palatini theories. This similarity is even more transparent when one realizes that the gravity Lagrangian  in the case of $f(\Rpal)$ theories is  $\LL_G=f(\Rpal)/2\kappa^2$ and in the Born-Infeld case, $\LL_{G}=\frac{\sqrt{|\Omega|}-\lambda}{\kappa^2\epsilon}$. Moreover, in the $f(\Rpal)$ theories, the conformal relation between the metrics can be seen as a particular case in which ${\Omega_\mu}^\nu=f_{\Rpal} {\delta_\mu}^\nu$. This allows us to express the field equations in the generic form
\begin{equation}\label{eq:FE-General}
{R^\mu}_\nu(h)=\frac{\kappa^2}{\sqrt{|\Omega|}}\left[\LL_G{\delta^\mu}_\nu+{T^\mu}_\nu\right] \ ,
\end{equation}
with ${\Omega_\mu}^\nu$ representing the relations (\ref{eq:Omega}), and the explicit dependence of ${\Omega^\mu}_\nu$  with the matter fields determined by the field equations of the specific theory. With formal manipulations, it is possible to show that this representation of the field equations in terms of the auxiliary metric $h_{\mu\nu}$ is indeed correct for large families of theories of gravity in which $\LL_G$ is just a functional of the inverse metric $g^{\mu\nu}$ and the Ricci tensor of an independent connection \cite{Jimenez:2015caa,Bazeia:2015zpa} (when torsion is set to zero at the end of the variation).  In vacuum configurations, the field equations recover GR plus an effective cosmological constant. \\

For convenience, we will use the generic equations (\ref{eq:FE-General}) to obtain formal expressions for the solutions of static, spherically symmetric configurations in which the stress-energy tensor possesses certain algebraic properties. These formal expressions will then be particularized to specific gravity plus matter models. \\

\section{Static, spherically symmetric solutions}\label{sec:generic}

In this section we will be concerned with stress-energy tensors with a specific algebraic structure, namely
\begin{equation}\label{eq:Tmn-generic}
{T^\mu}_\nu=\left(\begin{array}{cc} T_+ \hat I_{2\times2} & \hat O \\ \hat O & T_- \hat I_{2\times2} \end{array}\right) \ ,
\end{equation}
where $T_\pm$ are some functions of the space-time coordinates, $\hat I_{2\times2}$ is the $2\times 2$ identity matrix, and $\hat O$ is the $2\times 2$ zero matrix. Examples of stress-energy tensors with this structure arise in the case of electric fields and also for certain anisotropic fluids. The extension to higher-dimensions is straightforward using similar notation (see for instance \cite{Bazeia:2015zpa,Bazeia:2015uia}). 

Given that the deformation matrix ${\Omega^\mu}_\nu$ will be determined by the stress-energy tensor, we may assume that it also has a similar algebraic structure, i.e., we can take 
\begin{equation}\label{eq:Om-generic}
{\Omega^\mu}_\nu=\left(\begin{array}{cc} \Omega_+ \hat I_{2\times2} & \hat O \\ \hat O & \Omega_- \hat I_{2\times2} \end{array}\right) \ ,
\end{equation}
where $\Omega_\pm$ are given functions that should be provided by the field equations of the specific model considered. This point has been verified in several models explicitly and, therefore, appears as a reasonable assumption to proceed in a formal manner. \\

With the above assumptions, we find that the field equations (\ref{eq:FE-General}) become
\begin{equation}\label{eq:Rmn-generic}
{R^\mu}_\nu(h)=\frac{\kappa^2}{\sqrt{|\Omega|}}\left(\begin{array}{cc} (\LL_G+T_+) \hat I_{2\times2} & \hat O \\ \hat O & (\LL_G+T_-) \hat I_{2\times2} \end{array}\right) \ . 
\end{equation}
Now we need to focus on the form of the left-hand side to proceed further. For static, spherically symmetric configurations, we can take the line element of the space-time metric $g_{\mu\nu}$ as
\begin{equation}\label{eq:gmn-generic}
ds^2=g_{ab}(x^0,x^1)dx^a dx^b +r^2(x^0,x^1)(d\theta^2+\sin^2\theta d\varphi^2) \ ,
\end{equation}
where $(x^0, x^1)$ represent the coordinates of the $2\times 2$ sector orthogonal to the 2-spheres. Analogously, one can define a  line element for the auxiliary metric $h_{\mu\nu}$ of the form 
\begin{equation}\label{eq:hmn-generic}
d\tilde{s}^2=h_{ab}(x^0,x^1)dx^a dx^b +\tilde{r}^2(x^0,x^1)(d\theta^2+\sin^2\theta d\varphi^2) \ .
\end{equation}
Using the generic relations (\ref{eq:Omega}) between $h_{\mu\nu}$ and $g_{\mu\nu}$ together with (\ref{eq:Om-generic}), one finds that 
\begin{eqnarray}
h_{ab}&=&\Omega_+ g_{ab} \\
\tilde{r}^2&=&\Omega_- r^2 \ . \label{eq:x2r2}
\end{eqnarray}
For static configurations, we further specify the form of $h_{\mu\nu}$ as follows:
\begin{equation}\label{eq:hmn-concrete}
d\tilde{s}^2=-A(x) e^{2\Phi(x)}dt^2+\frac{1}{A(x)}dx^2+\tilde{r}^2(x)(d\theta^2+\sin^2\theta d\varphi^2) \ .
\end{equation}
Computing the Ricci tensor associated to this line element, one finds the following relations:
\begin{eqnarray}
{R_t}^t(h)&=& {R_x}^x(h)+\frac{4}{\tilde{r}}\left(\tilde{r}_{xx}-\Phi_x \tilde{r}_x\right) \\
{R_\theta}^\theta(h)&=& \frac{1}{\tilde{r}^2}\left[1-A \tilde{r}_x^2-\tilde{r} A \left(\tilde{r}_{xx}+\tilde{r}_x\left\{\frac{A_x}{A}+\Phi_x\right\}\right)\right] \ . 
\end{eqnarray}
Given that the right-hand side of (\ref{eq:Rmn-generic}) implies that ${R_t}^t= {R_x}^x$, it follows that $\left(\tilde{r}_{xx}-\Phi_x \tilde{r}_x\right)=0$. This equation allows us to take $\Phi(x)\to 0$ and $\tilde{r}\to x$, without loss of generality, and write the line element (\ref{eq:hmn-concrete}) in the form
\begin{equation}\label{eq:hmn-concrete1}
d\tilde{s}^2=-A(x)dt^2+\frac{1}{A(x)}dx^2+x^2(d\theta^2+\sin^2\theta d\varphi^2) \ .
\end{equation}
As a result, ${R_\theta}^\theta$ gets simplified as 
\begin{equation}
{R_\theta}^\theta(h)= \frac{1}{x^2}(1-A -x A_x) \ .
\end{equation}
It is now useful to insert the Ansatz 
\begin{equation}
A(x)=1-\frac{2M(x)}{x} \ ,
\end{equation}
which in combination with the right-hand side of (\ref{eq:Rmn-generic}) leads to the general expression
\begin{equation}\label{eq:Mx-general}
\frac{2M_x}{x^2}=\frac{\kappa^2}{\sqrt{|\Omega|}} (\LL_G+T_-) \ .
\end{equation}
Given that we are dealing with a static, spherically symmetric space-time, the functions appearing in the right-hand side of this equation are just functions of $x$ (or of $r(x)$). Therefore, by integrating this first-order equation, the  geometry will be completely determined. In practice, however, one still needs to find the explicit relation between the area functions $r^2(x)$ and $x^2$, which is specified by Eq.(\ref{eq:x2r2}). Recall, in this sense, that $\tilde{r}(x)\equiv x$ implies that $x^2=\Omega_- r^2$ and that, in general, $\Omega_-$ will be a function of $r$. This point will become clear when we consider explicit examples. \\

In the examples that we will consider below, the functions $\Omega_\pm$ depend on $x$ via $r(x)$. For this reason, it is convenient to express Eq. (\ref{eq:Mx-general}) in terms of the derivative with respect to $r$. This is immediate by just noting that $x^2=\Omega_- r^2$ implies
\begin{equation}
\frac{dr}{dx}=\frac{1}{\Omega_- ^{1/2}\left[1+\frac{1}{2}\frac{\Omega_{-,r}}{\Omega_-}\right]} \ .
\end{equation} 
The resulting expression for $M_r$ is thus 
\begin{equation}\label{eq:Mr-general}
M_r=\frac{\kappa^2\Omega_-^{1/2}}{2\Omega_+} (\LL_G+T_-) r^2\left[1+\frac{r}{2}\frac{\Omega_{-,r}}{\Omega_-}\right] \ .
\end{equation}
By integrating this equation, the space-time line element (defined by the metric $g_{\mu\nu}$) becomes 
\begin{equation}
ds^2=-\frac{A(x)}{\Omega_+}dt^2+\frac{1}{A(x)\Omega_+}dx^2+r^2(x)(d\theta^2+\sin^2\theta d\varphi^2) \ .
\end{equation} 
 In the next two sections we consider explicit examples that give concrete form to the above formulas.

\section{Solutions in Born-Infeld gravity. }\label{sec:BI}

Let us consider the coupling of the Born-Infeld gravity model to a spherically symmetric, static electric field defined by the action $S_M=-\frac{1}{16\pi}\int d^4x \sqrt{-g}F_{\mu\nu}F^{\mu\nu}$, being $F_{\mu\nu}$ the electromagnetic field strength tensor. For this matter source, the stress energy tensor can be written as
\begin{equation}\label{eq:Tmn-Maxwell}
{T_\mu}^\nu=\frac{q^2}{8\pi r^4}\left(\begin{array}{cc} -\hat I_{2\times2} & \hat O \\ \hat O & +\hat I_{2\times2} \end{array}\right) \ ,
\end{equation}
where $q$ represents the electric charge.  Inserting this expression in (\ref{eq:BI-h}), one finds that the components of ${\Omega^\mu}_\nu$ are just 
\begin{equation}\label{eq:Ompm0}
\Omega_\pm= 1\mp \frac{\epsilon \kappa^2 q^2}{8\pi r^4} \ . 
\end{equation}
Now we make a specific choice for the parameter $\epsilon$. Given that it has dimensions of squared length, we take $\epsilon=-2l_\epsilon ^2$, where $l_\epsilon$ represents some characteristic length scale. The sign of $\epsilon$ and the factor 2 have been chosen in such a way that the resulting solutions are identical to those found in the quadratic theory\footnote{From an algebraic point of view, it is much easier to deal with the Born-Infeld model \cite{Olmo:2013gqa} than with the above quadratic theory \cite{or12a}, though from an effective field theory approach it is easier to motivate the latter. For this reason we analyzed the field equations of the Born-Infeld model but restrict the discussion of solutions to those with more interest in the quadratic theory. We note that the sign in front of $l_\epsilon^2$ in (\ref{eq:f(R,Q)}) has been chosen in such a way that cosmological models with perfect fluids yield regular, bouncing solutions in both isotropic and anisotropic scenarios \cite{Barragan:2010qb}.}  
\begin{equation}\label{eq:f(R,Q)}
S=\frac{1}{2\kappa^2}\int d^4x \sqrt{-g}\left[R+l_\epsilon^2(a R^2+R_{\mu\nu}R^{\mu\nu})\right]-\frac{1}{16\pi}\int d^4x \sqrt{-g}F_{\mu\nu}F^{\mu\nu} \ .
\end{equation}
This is a curious property of the Born-Infeld and quadratic gravity theories that occurs in four space-time dimensions with stress-energy tensors of the form (\ref{eq:Tmn-generic}). With this choice, we can introduce a dimensionless variable $z=r/r_c$ such that $r_c^4\equiv l_\epsilon^2 r_q^2$, with $r_q^2\equiv \kappa^2 q^2/4\pi$, which turns 
(\ref{eq:Ompm0}) into 
\begin{equation}\label{eq:Ompm1}
\Omega_\pm= 1\pm \frac{1}{z^4} \ . 
\end{equation}
We can now use Eq.(\ref{eq:x2r2}), recalling that $\tilde{r}=x$, to find that
\begin{equation}\label{eq:r(x)BI}
r^2=\frac{x^2+\sqrt{x^4+4 r_c^4}}{2} \ .
\end{equation}
This relation puts forward that the area of the $2-$spheres has a minimum of magnitud $A_c=4\pi r_c^2$ at $x=0$. In other words, the sector $r<r_c$ is excluded from the range of values of the area function  $A=4\pi r^2(x)$. \\
\begin{figure}[h]
\begin{center}
\includegraphics[width=0.75\textwidth]{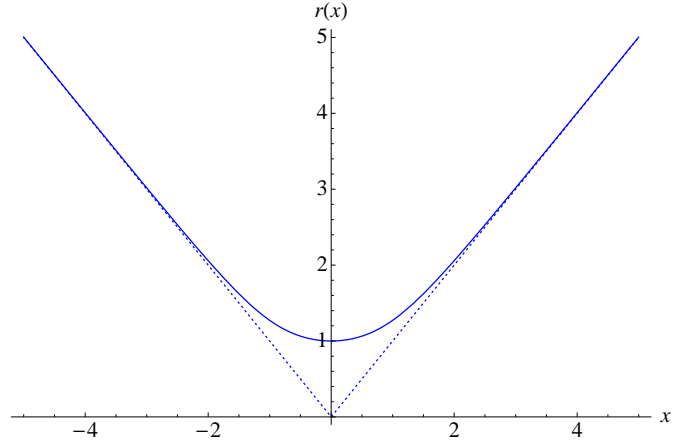}
\end{center}
\caption{ Representation of $r(x)$ (solid curve), defined in (\ref{eq:r(x)BI}), as a function of the radial coordinate $x$  in units of the scale $r_c$. The dotted lines represent the function $|x|$.   }\label{fig:r(x)BI}
\end{figure}

The mass function determined by Eq.(\ref{eq:Mr-general}) has a constant contribution and a term that comes from integrating over the electric field. The constant piece is identified with the Schwarzschild mass and will be denoted as $M_0$. To simplify the analysis, it is convenient to parametrize the mass function as follows:
\begin{equation}\label{eq:M}
M(r)=M_0(1+\delta_1 G(z)) \ ,
\end{equation} 
where $\delta_1$ is a dimensionless constant and $G(z)$ encodes the contribution of the electric field. Inserting this form of $M(r)$ in (\ref{eq:Mr-general}), one finds
\begin{equation}\label{eq:GzBI}
G_z=\frac{1}{z^4}\frac{(1+z^4)}{\sqrt{z^4-1}} \ ,  
\end{equation}
and 
\begin{equation}\label{eq:d1BI}
\delta_1=\frac{r_c^3}{2r_S l_\epsilon^2}=\frac{1}{2r_S }\sqrt{\frac{r_q^3}{ l_\epsilon}} \ ,
\end{equation}
where $r_S\equiv 2M_0$ denotes the Schwarzschild radius. The integration of $G_z$ is immediate and yields an infinite power series expansion of the form \cite{or12a}
\begin{equation}\label{eq:G(z)BI}
G(z)=-\frac{1}{\delta_c}+\frac{1}{2}\sqrt{z^4-1}\left[f_{3/4}(z)+f_{7/4}(z)\right] \ ,
\end{equation}
where $f_\lambda(z)={_2}F_1 [\frac{1}{2},\lambda,\frac{3}{2},1-z^4]$ is a hypergeometric function, and $\delta_c\approx 0.572069$ is a constant. Having obtained explicit solutions for $r^2(x)$ and $G(z)$, the space-time metric is completely specified.

\subsection{Properties and interpretation of the solutions}

One can verify from (\ref{eq:GzBI}) that for $z\gg 1$, $G(z)\approx -1/z$  yields the expected Reissner-Nordstr\"{o}m solution of GR, with $\Omega_\pm \approx 1$, $r^2(x)\approx x^2$, and
\begin{equation} \label{eq:RNsolution}
A(x)\approx 1-\frac{r_S}{ r  }+\frac{r_q^2}{2r^2}+O\left(\frac{r_c^4}{r^4}\right) \ .
\end{equation}
From this expression one readily verifies that the typical configurations in terms of horizons found for Reissner-Nordstr\"{o}m black holes also arise here, at least when the location of the horizon is much bigger than the scale $r_c$ \cite{or12a}. This occurs, in particular, when the charge-to-mass ratio $\delta_1$ is greater than $\delta_c$. We will refer to these configurations as RN-like.  When $\delta_1<\delta_c$, the solutions only have one horizon, like the Scharzschild black hole (Schwarzschild-like from now on). In some sense, the case $\delta_1<\delta_c$ describes the limit in which the charge is much smaller than the mass. When $\delta_1=\delta_c$, one finds a richer structure: depending on the number of charges, one can have one horizon, like in Schwarzschild, or have no horizons. More details on this will be given later. \\

It is apparent from (\ref{eq:GzBI}) and (\ref{eq:G(z)BI}) that the variable $z\equiv r/r_c$ can not become smaller than unity. This is consistent with (\ref{eq:r(x)BI}) and tells us that something relevant occurs at $r=r_c$ (or $z=1$ or $x=0$). Some information in this direction can already be extracted from the action that defines the theory. The fact that we are considering the combination of gravity with an electric field without sources means that our theory does not know about the existence of {\it sources} for the electric field. In GR, the   Reissner-Nordstr\"{o}m solution is derived under similar assumptions, and one considers that the solution is only valid outside of the sources, which are supposed to be somehow concentrated at the origin. This picture, however, is not completely satisfactory, and a precise description of the sources is still an open question (see chapter 8 of \cite{Ortin} for details). In our case, the combination of a minimum area for the two-spheres of the spherical sector together with the existence of an electric flux without sources points towards the notions of {\it geon} \cite{Wheeler:1955zz} and {\it wormhole} \cite{Misner:1957mt} suggested by J.A. Wheeler and C.W. Misner in the decade of 1950.  

It is well-known that an electric field flowing through a hole in the topology (wormhole) can generate a charge which, from all perspectives, acts exactly in the same way as point charges. Wormholes are characterized by having a minimum area, which defines their throat \cite{Visser:1995cc}. The Born-Infeld theory combined with a free Maxwell field considered here, therefore, is yielding self-gravitating wormhole solutions for which there is no need to consider additional sources \cite{Lobo:2013prg}. 

One should now note that in the derivation of the field equations we used a radial variable $x$ which was different from $r(x)$. The reason for this is that $r$ can only be used as a coordinate in those intervals in which it is a monotonic function of $x$ \cite{Stephani:2003ika}, and $r(x)$ has a minimum at the wormhole throat ($x=0$). Consistency of our {\it model of gravity plus electric field without sources} together with this behavior in the radial function implies the existence of a wormhole, in such a way that the range of $x$ is the whole real line (from $-\infty$ to $+\infty$). The theory is thus describing a spherically symmetric electric field which flows from one universe into another through a wormhole located at $x=0$  \cite{Lobo:2013prg}. On one of the sides, the electric field lines point in the direction of increasing area thus defining a positive charge. On the other side, the electric field points into the direction of decreasing area, defining in this way a negative charge. This type of configuration is similar to that envisioned by Einstein and Rosen \cite{Einstein:1935tc} when they used the Schwarzschild geometry to build a geometric model of elementary particles. A clear advantage of our model is that the wormhole structure arises naturally from the field equations and, therefore, one needs not follow a cut-and-paste strategy gluing together two exterior Schwarzschild geometries through the horizon to build the bridge that represents the particle in the Einstein-Rosen model. Moreover, a simple electric field has been able to generate a wormhole. This contrasts with the typical situation in GR, where wormholes supported by electric fields (linear like Maxwell's or nonlinear) are not possible \cite{Arellano:2006np}, being necessary exotic energy sources that violate the energy conditions \cite{Lobo:2007zb,Visser:1995cc}. \\

Having established the wormhole nature of our solutions, one should re-think the meaning of the classification given above regarding event horizons. What  we called Schwarzschild-like actually represents a wormhole with one horizon located somewhere on the $x>0$ side of the $x-$axis and another horizon symmetrical with this one but on the $x<0$ side. 
The RN-like configurations may have up to two horizons on each side of the $x-$axis. In the case with $\delta_1=\delta_c$, depending on the amount of electric charge (which is a measure of the intensity of the electric flux), we can have Schwarzschild-like configurations (one horizon on each side of the axis), a case in which the two horizons converge at $x=0$, and a horizonless family of (traversable) wormholes. This classification follows from a numerical study of the solutions of the equation $g_{tt}=-A/\Omega_+=0$ (see \cite{or12a} for details). \\

An analytical discussion of the behavior near the wormhole throat is possible and useful.  In fact, defining the number of charges as $N_q=q/e$, where $e$ is the proton charge, we have
\begin{equation}\label{eq:gtt_expansion}
\lim_{r\to r_c} g_{tt} \approx  \frac{l_P}{2l_\epsilon}\frac{N_q}{N_c}\left[-\frac{\left(\delta _1-\delta _c\right) }{2\delta _1 \delta _c }\sqrt{\frac{r_c}{ r-r_c} }+\left(1-\frac{l_\epsilon}{l_P}\frac{N_c}{N_q}\right)+ O\left(\sqrt{r-r_c}\right) \right]\ ,
\end{equation}
where, for convenience, we have introduced the Planck length $l_P=\sqrt{\hbar G/c^3}$ and $N_c\equiv \sqrt{2/\alpha_{em}}\approx 16.55$, with  $\alpha_{em}$ representing the electromagnetic fine structure constant. This expression puts forward that the metric is finite at $r=r_c$ only for $\delta_1=\delta_c$,  diverging otherwise.  By direct computation one can verify that curvature scalars generically diverge at $r=r_c$ except for those solutions with $\delta_1=\delta_c$, where constant scalars are obtained. For this regular case, Eq. (\ref{eq:gtt_expansion}) also shows that the wormhole is hidden behind an event horizon if the sign of $\left(1-\frac{l_\epsilon}{l_P}\frac{N_c}{N_q}\right)$ is positive, because then $g_{tt}>0$ near the throat. 

If we take $l_\epsilon=l_P$, i.e., if the characteristic length scale of the gravity sector coincides with the Planck scale, then the event horizon for the regular solutions exists if $N_q>N_c$. For smaller values of the charge, $N_q\leq 16.55$, the horizon disappears and we are left with a regular horizonless object which could be interpreted as a black hole remnant. The existence of this type of solutions is interesting for theoretical as well as for astrophysical reasons.  Theoretically, the existence of regular remnants could have important implications for the quantum information loss in the process of black hole evaporation \cite{Fabbri:2005mw}. From an astrophysical perspective, the existence of remnants could justify the lack of observational evidence for black hole explosions. Moreover, solutions of this type could contribute to the so-called dark matter in the form of very massive neutral atoms \cite{Lobo:2013prg}. In fact, from the charge-to-mass constraint $\delta_1=\delta_c$, one finds that the mass of these solutions is completely determined by their electric charge according to the formula
\begin{equation}\label{eq:M0BI}
M_0= n_{BI} m_P \left(\frac{N_q}{N_c}\right)^{3/2}\left(\frac{l_P}{l_\epsilon}\right)^{1/2} \ ,
\end{equation}
where $n_{BI}=\pi^{3/2}/(3\Gamma[3/4]^2)\approx 1.23605$ is a number that also arises in the determination of the total electrostatic energy of a point charge in the Born-Infeld theory of electrodynamics\footnote{In fact, using a notation similar to ours, in the Born-Infeld electromagnetic theory , whose Lagrangian is $\LL_{BI}=\beta^2\left(\sqrt{-|\eta_{\mu\nu}+\beta^{-1}F_{\mu\nu}|}-\sqrt{-|\eta_{\mu\nu}|}\right)$, one finds that the total electrostatic energy of a point particle is $
\mathcal{E}_{BI}= \sqrt{2}n_{BI} m_P c^2\left(\frac{N_q}{N_c}\right)^{3/2}\left(\frac{l_P}{l_\beta}\right)^{1/2}$, 
where $l_\beta^2\equiv (4\pi/\kappa^2c \beta^2)$ is a length scale associated to the $\beta$ parameter of the theory.} (formulated in flat Minkowski space-time). With the mass formula (\ref{eq:M0BI}), one can verify that Hawking's original predictions regarding the mass and charge spectrum of  primordial black holes \cite{Hawking:1971ei} formed in the early universe are in consonance with our results. He found that collapsed objects of order the Planck mass and above and with up to $\pm 30$ electron charges could have been formed by large density fluctuations. It is typically argued that the existence of a quantum instability due to the horizon would make the lightest primordial black holes decay and evaporate. With the above explicit results, it is apparent that new mechanisms could lead to the formation of stable remnants which could survive until our times. \\ 

As a curiosity, from (\ref{eq:M0BI}) one also finds that a solar mass black hole (with $\sim 10^{57}$ protons) of this type would require only $N_q\sim 3 \times 10^{26}$ charges (or $\sim 484$ moles) to make the metric and all curvature scalars regular at the origin. Moreover, the external horizon of such an object would almost coincide with the Schwarzschild radius predicted by GR, making these objects astrophysically identical to those found in GR. This amount of charge  certainly allows us to get rid of a number of important problems at a very low price. However, one should recall that  (\ref{eq:M0BI}) is only strictly valid for the $\delta_1=\delta_c$ configuration, which suggests that only fine tuned configurations would be satisfactory. This raises a natural question: given that for $\delta_1=\delta_c$ the  geometry is completely regular and that infinitesimal deviations from this relation imply the development of curvature divergences and infinities in the metric, what happens to geodesics? In the $\delta_1=\delta_c$ case we expect geodesics to be complete, as there is no reason to expect any pathological behavior that limits their extendibility at or near the wormhole throat. What happens to them when $\delta_1\neq\delta_c$? Answering this question will provide us with useful information on the relation between curvature divergences  and the existence of observers. In other words, this model offers us a good opportunity to better understand the correlation existing in GR between curvature divergences and geodesic incompleteness. We will resume this discussion later on, when we consider the geodesic equation in Sec. \ref{sec:geodesics}.

\section{Solutions in $f(\Rpal)=\Rpal-\lambda \Rpal^2$. } \label{sec:f(R)}

In Sec. \ref{sec:EOM} we discussed the field equations of the Palatini version of $f(R)$ theories. Now we would like to find nontrivial black hole solutions and study their properties to see how their geodesic structure compares with that provided by GR.
A natural procedure would be to consider the coupling of an electric field as we did in the previous section in the case of Born-Infeld gravity. However, given that the stress-energy tensor of Maxwell's electrodynamics is traceless and that the modified dynamics of Palatini $f(R)$ theories depends crucially on nonlinear functions of this trace, we find that electrovacuum solutions in these theories are identical to those found in GR with a cosmological constant. Thus, in order to explore new physics, we need to consider matter sources whose stress-energy tensor has a non-zero trace. \\

To proceed, we consider a generic anisotropic fluid with stress-energy tensor of the form \cite{Harko:2013aya,Shaikh:2015oha}
\begin{equation}
{T_\mu}^\nu =\left(\begin{array}{cccc} -\rho & 0 & 0 & 0 \\ 0 & P_r & 0 & 0 \\ 0 & 0 & P_\theta & 0 \\ 0 & 0 & 0 & P_\varphi \end{array}\right)
\end{equation}
and set $P_r=-\rho$ and $P_\theta=P_\varphi=K(\rho)$, where $K(\rho)$ is some function of the fluid density, such that our fluid has the same structure as the generic stress-energy tensor considered in Sec. \ref{sec:generic}
\begin{equation}
{T_\mu}^{\nu}={diag}[-\rho,-\rho, K(\rho),K(\rho)] \ .
\end{equation}
 It is worth noting that this structure of the stress-energy tensor allows us to see it as corresponding to a non-linear theory of electrodynamics \cite{Olmo:2015axa}. In fact, for a theory where the electromagnetic Lagrangian goes from $X=-\frac{1}{2}F_{\mu\nu} F^{\mu\nu}$ to $\varphi(X)$, the stress-energy tensor becomes
\begin{equation}
{T_\mu}^{\nu}=\frac{1}{8\pi}{diag}[\varphi-2X\varphi_X, \varphi-2X\varphi_X, \varphi,\varphi] \ .
\end{equation}
We can thus establish the correspondences $-8\pi \rho= (\varphi-2X\varphi_X)$ and $K(\rho)=\varphi(X)$, which allow to solve for $\varphi(X)$ once a function $K(\rho)$ is specified. \\

Considering the fluid representation, the conservation equation $\nabla_\mu {T^\mu}_\nu=0$ for a line element of the form $ds^2=-C(x)dt^2+B^{-1}(x) dx^2+r^2(x)(d\theta^2+\sin^2\theta d\varphi^2)$ leads to the relation $\rho_x+2[\rho+K(\rho)]r_x/r=0$. This expression can be readily integrated to obtain a formal relation between $\rho(x)$ and $r(x)$ given by
\begin{equation}
r^2(x)=r_0^2 \exp\left[{-\int^\rho \frac{d\tilde{\rho}}{\tilde{\rho}+K(\tilde{\rho})}}\right] \ ,
\end{equation}
where $r_0$ is an integration constant with dimensions of length. In order to simplify our discussion, we shall restrict ourselves to the case $K(\rho)=\alpha \rho+\beta \rho^2$, where $\alpha$ is a dimensionless constant and $\beta$ has dimensions of inverse density. This example yields analytical solutions and covers a number of interesting cases. In particular, one finds that the relation between $\rho(x)$ and $r(x)$ turns into
\begin{equation}
\rho(r)=\frac{(1+\alpha) \rho_0}{\left(\frac{r}{r_0}\right)^{2(1+\alpha)}-\beta \rho_0} \ .
\end{equation}
 One readily verifies that when $\alpha=1$ and $\beta=0$, this fluid has the same stress-energy tensor as the Maxwell electric field (\ref{eq:Tmn-Maxwell}), with $\rho_0 r_0^4=q^2/8\pi$. The inclusion of the parameters $\alpha$ and $\beta$ allows to generate a non-zero trace in the stress energy tensor. The case with $\beta=0$ and $0<\alpha<1$ was studied in detail in \cite{Olmo:2015axa}. Here we shall take $\alpha=1$ and focus on the case $\beta<0$ (a more exhaustive discussion will be presented elsewhere \cite{Bejarano}). This family of models rapidly recovers the usual RN solution away from the center but regularizes the energy density, which is everywhere finite and bounded above by  $\rho_{m}=\frac{(1+\alpha)}{|\beta|}$. We note that the effect of the parameter $\beta>0$ is to shift the location of the divergence in the density from $r=0$ to $(|\beta|\rho_0)^{1/(2+2\alpha)} r_0$. With our choice of negative $\beta$, we regularize the divergence of the matter sector. \\

To proceed, we set $\alpha=1$, $\beta=-\tilde{\beta}/\rho_0$, and introduce a dimensionless variable $z^4=r^4/\tilde{\beta} r_0^4$, in such a way that the density is now given by 
\begin{equation}
\rho=\frac{\rho_m}{1+z^4}  \ .
\end{equation}
Using the trace equation (\ref{eq:tracefR}) and the quadratic model $f=\Rpal-\lambda \Rpal^2$, one readily finds that $\Rpal=-\kappa^2T$, which is the same linear relation as in GR (this is just an accident of the quadratic model in four dimensions). We thus find that the function $f_{\Rpal}$ takes the simple form
\begin{equation}
f_{\Rpal}=1-\frac{\gamma}{(1+z^4)^2} \ ,
\end{equation}
where $\gamma\equiv \rho_m/\rho_\lambda$ and $\rho_\lambda\equiv 1/8\kappa^2\lambda$. \\

\begin{figure}[h]
\begin{center}
\includegraphics[width=0.75\textwidth]{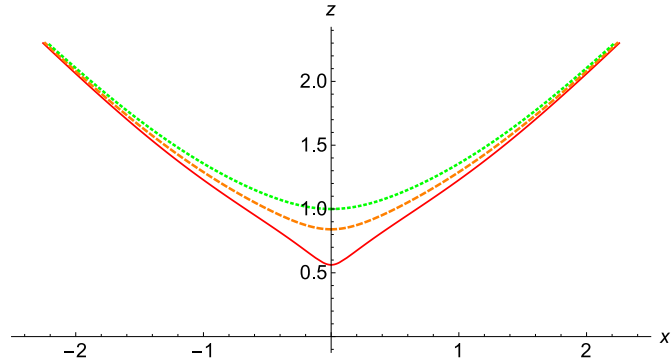}
\end{center}
\caption{ Representation of $z(x)$ (solid curve) as a function of the radial coordinate $x$  (in units of the scale $r_c=|\tilde{\beta}|^{1/4}r_0$) for different values of the parameter $\gamma$. The solid (red) curve corresponds to $\gamma=1.1$, the dashed (orange) curve is $\gamma=1.5$, and $\gamma=2$ is the dotted (green) one.}\label{fig:BouncefR}
\end{figure}

Following the same approach as in the Born-Infeld gravity theory studied above, we find that parametrizing the mass function as $M(r)=M_0(1+\delta_1 G(z))$ leads to 
\begin{eqnarray}\label{eq:GzfR}
G_z&=&\frac{z^2}{(1+z^4)f_{\Rpal}^{3/2}}\left(1-\frac{\gamma}{(1+z^4)^3}\right)\left(1-\frac{\gamma(1-3z^4)}{(1+z^4)^3}\right) \\ 
\delta_1&\equiv& \frac{\kappa^2\rho_m (r_0\tilde{\beta}^{\frac{1}{4}})^{3}}{r_S}
\end{eqnarray}                
The function $G(z)$ can be obtained easily in terms of power series expansions and the solutions are classified in two types, depending on the value of the parameter $\gamma$. If $\gamma>1$ then $z$ is bounded from below, $z\ge z_c$, with $z_c^4=\gamma^{1/2}-1$ representing the location where $f_{\Rpal}=0$.  
At that point, the function $G_z$ diverges, as can be easily understood from the expression (\ref{eq:GzfR}), which has a term $f_{\Rpal}^{3/2}$ in the denominator. The lower bound on $z$ signals the presence of a wormhole, in much the same way as we already observed in the case of Born-Infeld gravity. This is confirmed by the relation between the radial functions $x$ and $z$ given by $x^2=f_{\Rpal} z^2$, which is plotted in Fig.\ref{fig:BouncefR}. Having this wormhole structure in mind, one finds that near $z_c$ we have $f_{\Rpal}\approx \frac{8z_c^3}{1+z_c^4}(z-z_c)$ and $G_z\approx C/(z-z_c)^{3/2}$, with $C>0$ a constant (whose explicit form can be computed but is not necessary). This leads to $\lim_{z\to z_c} G(z)\approx -2C/\sqrt{z-z_c}$. \\

 It is obvious that for $0<\gamma<1$ there are no real solutions for $z_c$. One finds that for that case, and also for $\gamma=1$, the range of $z$ is comprised between $0$ and $\infty$, which implies that there is no wormhole, $G_z$ is finite everywhere, and $G(z)$ tends to a constant as $z\to 0$. In fact, near $z=0$ we can approximate $G(z)\approx -\frac{1}{\delta_c^{(\gamma)}}+(1-\gamma)^{1/2}z^3/3 +\frac{(7\gamma-1)}{\sqrt{1-\gamma}}{z^7}/{7}+O(z^{11})$, where $\delta_c^{(\gamma)}$ is a constant. The case $\gamma=1$ admits an analytical solution in terms of special functions and its series expansion must be considered separately, yielding $G(z)\approx -1/\delta_c^{(1)}+\frac{9 z^5}{5 \sqrt{2}}-\frac{13 z^9}{4 \sqrt{2}}+O(z^{13})$.  One can easily verify that for $z\gg 1$ (\ref{eq:GzfR}) rapidly converges to the GR prediction $G_z\approx 1/z^2$ regardless of the value of $\gamma$. \\

Let us now discuss the geometry near the center in the two cases distinguished above in terms of $\gamma$. Consider first the wormhole case, $\gamma>1$, for which $\lim_{z\to z_c} f_{\Rpal}\approx \frac{8z_c^3}{1+z_c^4}(z-z_c)$ and  $\lim_{z\to z_c} G(z)\approx -2C/\sqrt{z-z_c}$. The area of the two spheres is determined by solving the relation $x^2=f_{\Rpal} r^2$. Denoting $r=z r_c$, $x=\tilde{x} r_c$, and $r_c=r_0 \tilde{\beta}^{1/4}$, one finds 
\begin{equation}
\tilde{x}\approx \sqrt{ \frac{8z_c^5}{1+z_c^4}}(z-z_c)^{1/2} \ ,
\end{equation}
which leads to 
\begin{equation}\label{eq:r2fR}
r^2(x)\approx r_c^ 2z_c^2+\frac{(1+z_c^ 4)}{4z_c^4}x^2 \ .
\end{equation}
This relation puts forward that the physical $2-$spheres have a minimum area at $x=0$, thus signaling the presence of a wormhole, as already advanced above. The $g_{tt}$ component of the metric can be written as
\begin{equation}
g_{tt}=-\frac{1}{f_{\Rpal}}\left(1-\frac{r_S (1+\delta_1 G(z))}{x}\right)\approx -\frac{\tilde{C}}{(z-z_c)^2} \ ,
\end{equation}
where $\tilde{C}$ is a positive constant whose explicit form is not relevant. It is clear that for this type of solutions the metric diverges at $z=z_c$. One can also verify that curvature scalars generically diverge on that surface.  We note that the properties of the solutions with $\gamma>1$ are shared by all those models in which $f_{\Rpal}$ has a simple pole at $z=z_c$. One can easily verify that if $f_R=b_0 (z-z_c)$, then the two spheres satisfy a relation like (\ref{eq:r2fR}) and the metric has a quadratic divergence at $z_c$. \\

When $0<\gamma\le 1$, the properties of the solutions largely depart from those observed in the case of  having a pole in $f_{\Rpal}$. Given that the function $f_{\Rpal}$ does not vanish in this case, we find that near the center $\tilde{x}\approx \sqrt{1-\gamma} \ z$. The $g_{tt}$ component of the metric then becomes
\begin{equation}
g_{tt}\approx-\frac{1}{(1-\gamma)}\left(1-\frac{r_S (\delta_c^{(\gamma)}-\delta_1)}{r_c \delta_c^{(\gamma)}\sqrt{1-\gamma} \ z}-\frac{r_S\delta_1}{2r_c}z^2\ldots\right) \ .
\end{equation}
This indicates that for the choice $\delta_1=\delta_1^{(\gamma)}$, the metric is regular everywhere. Curvature scalars, however, do have divergences. For $\gamma=1$, the above expression must be replaced by 
\begin{equation}
g_{tt}\approx\frac{{r_S}}{2r_c \sqrt{2} z^7}-\frac{1}{2 z^4}+O(z^{-3}) \ .
\end{equation}
We note that the case $\gamma\to 0$ yields the limit in which this anistropic fluid is coupled to GR. One can verify that the behavior of the solutions with $0<\gamma\le 1$ near the origin is similar to that of models of nonlinear electrodynamics coupled to GR \cite{Olmo:2011ja,NED1,NED2,NED3,NED4,AB1,AB2,AB3,NED5,NED6,NED7,NED8,NED9,NED10,NED11}.

\section{Geodesics.}\label{sec:geodesics}

The modified gravitational dynamics generated by the models considered in the previous sections has an impact on the space-time metric $g_{\mu\nu}$ and, consequently, on its associated geodesics. Since we are interested in determining whether the space-times derived above are geodesically complete or not, in this section we solve the geodesic equation and explore their behavior in those regions where GR typically yields incomplete paths. \\

The geodesics of a given connection $\Gamma^\mu_{\alpha\beta}$ are determined by the equation 
\begin{equation}\label{eq:geodesics}
\frac{d^2x^\mu}{d\lambda^2}+\Gamma^\mu_{\alpha\beta}\frac{dx^\alpha}{d\lambda}\frac{dx^\beta}{d\lambda}=0 \ . 
\end{equation}
Here we will focus on the geodesics of the metric $g_{\mu\nu}$, which are the ones that matter fields can see according to the Einstein equivalence principle.  We thus take $\Gamma^\mu_{\alpha\beta}$ as defined in (\ref{eq:LC}). In order to solve these equations, we introduce a Hamiltonian approach that simplifies the analysis. To proceed, we first note that (\ref{eq:geodesics}) can be derived from an action of the form \cite{Chandra}
\begin{equation}
S=\frac{1}{2}\int d\lambda g_{\mu\nu} \frac{dx^\mu}{d\lambda}\frac{dx^\nu}{d\lambda} \ ,
\end{equation}
which for a line element like $ds^2=-C(x)dt^2+B^{-1}(x)dx^2+r^2(x)d\Omega^2$ becomes
 \begin{equation}
S=\frac{1}{2}\int d\lambda \left[-C(x)\dot{t}^2+\frac{1}{B(x)}\dot{x}^2+r^2(x)\dot{\theta}^2+r^2(x)\sin^2\theta \dot{\varphi}^2\right] \ .
\end{equation}
From this representation, one easily verifies that the momenta associated to the variables $(t,x,\theta,\varphi)$ are
\begin{eqnarray}
P_t&=&-\frac{\partial L}{\partial \dot{t}}=\dot{t} C(x) \\ 
P_x&=&\frac{\partial L}{\partial \dot{x}}=\dot{x}/B(x) \\ 
P_\theta&=&\frac{\partial L}{\partial \dot{\theta}}=r^2(x)\dot{\theta} \\ 
P_\varphi &=&\frac{\partial L}{\partial \dot{\varphi}}=r^2(x)\sin^2\theta \dot{\varphi} \ . 
\end{eqnarray}
With these momenta one finds that the Hamiltonian $H=-P_t \dot{t}+P_x \dot{x}+P_\theta \dot{\theta}+P_\varphi \dot{\varphi}-L$ coincides with the Lagrangian (due to the absence of potential terms) and can be written as
\begin{equation}
H=\frac{1}{2}g^{\mu\nu}(x)P_\mu P_\nu \ .
\end{equation}
The geodesic equations can thus be written as 
\begin{eqnarray}
\dot{x}^\mu&=& \frac{\partial H}{\partial P_\mu}=g^{\mu\nu}P_\nu \\
\dot{P}_\mu&=& -\frac{\partial H}{\partial x^\mu}=-\frac{1}{2}(\partial_\mu g^{\alpha\beta})P_\alpha P_\beta 
\end{eqnarray}
From these equations one readily sees that $P_t$ and $P_\varphi$ are constants of the motion, as $\dot{P}_t=0=\dot{P}_\varphi$. These equations also imply that $dH/d\lambda=0$, showing that $H$ is another conserved quantity.  
We thus have
\begin{eqnarray}
P_t&=&\left(\frac{dt}{d\lambda}\right)C(x)=E \\ 
P_\varphi &=&\left(\frac{d\varphi}{d\lambda}\right)r^2(x)\sin^2\theta=L \\ 
2H&=& -\frac{P_t^2}{C(x)}+B(x)P_x^2+\frac{P_\theta^2}{r^2(x)}+\frac{P_\varphi^2}{r^2(x)\sin^2\theta}=-\frac{E^2}{C(x)}+\frac{\dot{x}^2}{B(x)}+\frac{L^2}{r^2} \ , \label{eq:2H}
\end{eqnarray} 
where in the last equality we have set $\theta=\pi/2$ without loss of generality (because the motion takes place on a plane). When $H\neq 0$,  a constant rescaling of the affine parameter $\lambda\to \lambda/\sqrt{|2H|}$ makes it clear that only the sign of $H$ is physically relevant. This sign allows to classify the geodesics in three families: those with $H>0$ (space-like), those with $H<0$ (time-like), and those with $H=0$ (null), which clarifies the meaning of this conserved quantity. Denoting $k\equiv 2H$ (with $k=1,0,-1$ corresponding to spatial, null, and time-like geodesics, respectively), Eq. (\ref{eq:2H}) can be recast as 
\begin{equation}\label{eq:geo_general}
\frac{C(x)}{B(x)}\left(\frac{dx}{d\lambda}\right)^2=E^2-C(x)\left(\frac{L^2}{r^2(x)}-k\right) \ ,
\end{equation}
which will be used to study the range of $\lambda$ in different scenarios. 

\subsection{Geodesics in GR}

Let us consider the Schwarzschild and Reissner-Nordstr\"{o}m solutions of GR, whose line element takes the form
\begin{equation}\label{eq:ds2GR}
ds^2=-C(r)^2dt^2+\frac{1}{C(r)}dr^2+r^2 d\Omega^2 \ ,
\end{equation}
with $C(r)=1-\frac{r_S}{r}+\frac{r_q^2}{2r^2}$, $r_S=2GM_0/c^2$, $r_q^2=\kappa^2 q^2/4\pi$ (for Schwarzschild, $r_q^2=0$), and $\kappa^2=8\pi G/c^4$. Given that here $C(r)=B(r)$, we find that (\ref{eq:geo_general}) turns into
\begin{equation}\label{eq:geoGR}
\left(\frac{dr}{d\lambda}\right)^2=E^2-C(r)\left(\frac{L^2}{r^2}-k\right) \ .
\end{equation}
This equation has the same structure as that of a particle with energy $\mathcal{E}=E^2$ in an effective one-dimensional potential of the form $V_{eff}(r)=C(r)\left(\frac{L^2}{r^2}-k\right) $, which facilitates its interpretation. \\

Let us consider first the uncharged (Schwarzschild) case. In this scenario, the function $C(r)$ becomes negative inside the horizon. As a result, the effective potential becomes an infinitely attractive well of the form $V_{eff}\approx -\frac{r_S}{r}\left(\frac{L^2}{r^2}-k\right)$, and the causal structure is such that all observers and light rays are forced to move in the direction of decreasing $r$ as time goes by.
This can be seen straightforwardly by just writing the line element (\ref{eq:ds2GR}) in ingoing Eddington-Finkelstein coordinates
 \begin{equation}\label{eq:ds2GRv}
ds^2=-C(r)^2dv^2+2dvdr+r^2 d\Omega^2 \ ,
\end{equation}
where $dv=dt+dr/C(r)$ now plays the role of time coordinate. Inside the event horizon, where $A(r)<0$, we see that 
 \begin{equation}
-2dvdr=-C(r)^2dv^2-ds^2+r^2 d\Omega^2 \ 
\end{equation}
implies that as time goes by ($dv>0$) we must have $dr<0$ for time-like and null trajectories ($ds^2\leq 0$). Thus, regardless of their point of origin, all physical observers and light rays will sooner or later end up at $r=0$. The precise evolution of the affine parameter near the center is determined by $dr/d\lambda\approx - \sqrt{r_S/r}$ for radial timelike geodesics ($L=0$) and by $dr/d\lambda\approx - \sqrt{r_SL^2/r^3}$ for timelike and null geodesics with $L\neq 0$. By integrating these expressions, we find $\lambda(r)=\lambda_0-\frac{2}{3}\sqrt{r^3/r_S}$ and $\lambda(r)=\lambda_0-\frac{2}{5}\sqrt{r^5/r_SL^2}$, respectively, where $\lambda_0$ represents the value of the affine parameter at $r=0$. Given that the affine parameter cannot be extended {\it beyond} the center, these geodesics are incomplete in the future. A similar analysis can be carried out in the {\it white hole} region of the Schwarzschild geometry, where all geodesics are outgoing ($dr>0$ with growing time). In that case, geodesics are incomplete in the past, i.e., they cannot be extended into $\lambda\to -\infty$.  This space-time, therefore, can be regarded as singular. \\

In the Reissner-Nordstr\"{o}m case, the situation is quite different from Schwazschild. As one approaches the center, the charge term dominates and $C(r)\sim \frac{r_q^2}{2r^2}>0$ implies that for time-like observers ($k=-1$) $dr/d\lambda$ in (\ref{eq:geoGR}) must vanish at some point before reaching $r=0$ regardless of the value of $L$. These observers, therefore, bounce before reaching the center due to the presence of an infinite potential barrier and continue their trip in the direction of growing $r$, having the possibility of getting into new asymptotically flat regions if horizons are present. Something similar happens also to light rays ($k=0$) with nonzero angular momentum $L$. However, for radial null geodesics ($k=0$ and $L=0$),  we find $r(\lambda)=\pm E (\lambda-\lambda_0)$, where the minus sign represents ingoing rays and the plus sign outgoing rays. Ingoing rays cannot be extended beyond $\lambda=\lambda_0$, whereas outgoing rays are {\it created} at some finite $\lambda$. Thus,  the Reissner-Nordstr\"{o}m geometry is incomplete as far as radial null geodesics are concerned.

\subsection{Geodesics in Born-Infeld gravity}

From our discussion of the spherically symmetric charged solutions found in Sec. \ref{sec:BI} for the Born-Infeld theory, it is clear that geodesics in that space-time are essentially the same as in GR as soon as one moves a few $r_c$ units away from the central wormhole \cite{Olmo:2015bya}. In fact, in Fig. \ref{fig:r(x)BI} one can readily see that $r(x)\approx x$ as soon as one reaches $|x|\approx 2r_c$. The $g_{tt}$ component of the metric also converges quickly to the GR prediction, as shown in (\ref{eq:RNsolution}), with corrections that decay rapidly as $\sim (r_c/r)^4$. We thus only need to focus on the behavior of geodesics near the wormhole to explore the impact of curvature divergences on their completeness. Recall, in this sense, that the different metric solutions could be classified according to whether the charge-to-mass ratio $\delta_1$, defined in (\ref{eq:d1BI}), was smaller, equal, or larger than the characteristic value  $\delta_c\approx 0.572069$ that arises in the electric field contribution to the mass function of Eq.(\ref{eq:G(z)BI}). The case $\delta_1=\delta_c$ was completely regular (no metric or curvature divergences \cite{or12a}), whereas $\delta_1<\delta_c$ (Schwarzschild-like) and $\delta_1>\delta_c$ (RN-like) had divergences at the wormhole throat, $x=0$ (or $r=r_c$ or $z=1$).\\

\begin{figure}[h]
\begin{center}
\includegraphics[width=0.75\textwidth]{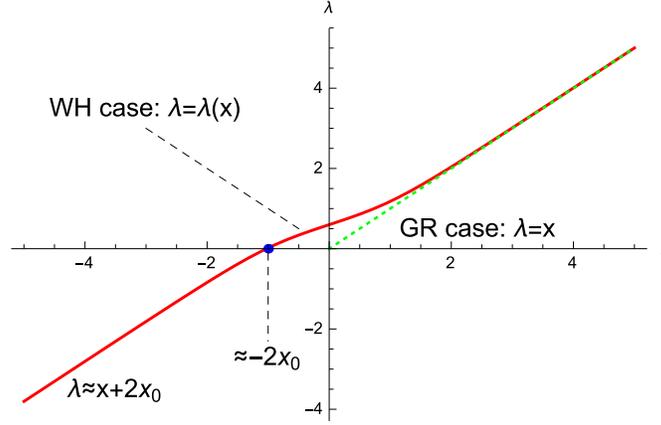}
\end{center}
\caption{Affine parameter $\lambda(x)$ as a function of the radial coordinate $x$ for radial null geodesics (outgoing in $x>0$). In the GR case (green dashed curve in the upper right quadrant), $\lambda=x$ is only defined for $x\ge 0$. For radial null geodesics in our wormhole spacetime (solid red curve), $\lambda(x)$ interpolates between the GR prediction and a shifted straight line $\lambda(x)\approx  x+2x_0$, with $x_0\approx 0.59907$. In this plot $E=1$ and the horizontal axis is measured in units of $r_c$. } \label{Fig:affine_nullradial}
\end{figure}

Using the identifications $C(x)=A(x)/\Omega_+$ and $B(x)=A(x)\Omega_+$ together with the expression for $r^2(x)$ found in (\ref{eq:r(x)BI}),  Eq. (\ref{eq:geo_general}) turns into
\begin{equation}\label{eq:geoBI}
\frac{1}{\Omega_+^2}\left(\frac{dx}{d\lambda}\right)^2=E^2-\frac{A(x)}{\Omega_+}\left(\frac{L^2}{r^2(x)}-k\right) \ .
\end{equation} 
For radial null geodesics ($L=0, \ k=0$), which are incomplete in both the Schwarzschild and RN solutions of GR, the above equation becomes independent of the function $A(x)$ and an exact solution can be found analytically. Using Eq. (\ref{eq:r(x)BI}), one finds that $dx/dr=\pm\Omega_+/\Omega_-^{1/2}$, with the minus sign corresponding to $x\le 0$. This turns (\ref{eq:geoBI}) into
\begin{equation}\label{eq:geoBI}
\frac{1}{\Omega_-}\left(\frac{dr}{d\lambda}\right)^2=E^2 \ , 
\end{equation}
which can be integrated to obtain
\begin{equation}\label{eq:nullradial2}
\pm E \cdot \lambda(x)=\left\{ \begin{array}{lr} {_{2}{F}}_1[-\frac{1}{4},\frac{1}{2},\frac{3}{4};\frac{r_c^4}{r^4}]  r & \ x\ge 0 \\
{ }\\
2x_0- {_{2}{F}}_1[-\frac{1}{4},\frac{1}{2},\frac{3}{4};\frac{r_c^4}{r^4}]  r &  \ x\le 0
\end{array} \right. \ ,
\end{equation}
where $_{2}F_1[a,b,c;y]$ is a hypergeometric function, $x_0={_{2}{F}}_1[-\frac{1}{4},\frac{1}{2},\frac{3}{4};1] =\frac{\sqrt{\pi}\Gamma[3/4]}{\Gamma[1/4]}\approx 0.59907$, and the $\pm$ sign corresponds to outgoing/ingoing null rays in the $x>0$ region. It should be noted that given that $dr/d\lambda$ is a continuous function, the solution (\ref{eq:nullradial2}) is unique. One can easily verify that as $x\to \infty$ the series expansion of (\ref{eq:nullradial2})  yields $\pm E\lambda(x) \approx r+O(r^{-3})\approx x$ and naturally recovers the GR behavior for large radii (see Fig.\ref{Fig:affine_nullradial}).  As $x\to -\infty$, we get $\pm E\lambda(x)\approx x+2x_0$, which also recovers the linear behavior of GR but shifted by a (negligible)  constant factor.

Given that the radial coordinate $x$ can naturally take negative values due to the wormhole structure, it follows that the affine parameter for radial null geodesics can  be extended over the whole real line. As a result, these geodesics are complete. This was expected for the regular case with $\delta_1=\delta_c$, for which the metric and all curvature scalars are finite everywhere, but was not obvious {\it a priori} for the other cases. Remarkably, the fact that this result is independent of the details of the function $A(x)$, which contains the information about $\delta_1$, confirms that radial null geodesics are complete for all our solutions. This puts forward that a space-time can be geodesically complete even when there exist divergences in the metric and/or in curvature scalars.  The wormhole has thus crucially contributed to allow the extendibility of the most critical geodesics of GR. \\

For nonradial and/or time-like geodesics, the discussion must take into account whether the geometry is Schwarzschild-like or RN-like. Considering the limit $x\to 0$, Eq.(\ref{eq:geoBI}) turns into 
\begin{eqnarray}\label{eq:geoBInear}
\frac{1}{4}\left(\frac{dx}{d\lambda}\right)^2&=&E^2-V_{eff}(x) \\
\label{eq:Vzero}
V_{eff}(x)&\approx& -\frac{a}{|x|}-b \ ,
\end{eqnarray}
with $a=\left(\kappa+\frac{L^2}{r_c^2}\right)\frac{  (\delta_c-\delta_1)}{2\delta_c \delta_2 }$, $b=\left(\kappa+\frac{L^2}{r_c^2}\right)\frac{(\delta_1-\delta_2) }{2 \delta _2}$, and $\delta_2\equiv \delta_1 \frac{N_c}{N_q}\frac{l_\epsilon}{l_P}$. From the above expression it is easy to see that in the RN-like configuration the coefficient $a$ is negative, thus implying that the right-hand side of (\ref{eq:geoBInear}) must vanish at some point before reaching the wormhole. The situation is thus analogous to that already observed in the case of GR, with $L\neq 0$ geodesics bouncing before reaching the center (or the wormhole in our case). In the Schwarzschild-like configurations, the effective potential represents an infinite attractive well with the possibility of having a maximum before reaching the throat. As a consequence,  all geodesics with energy above that maximum hit the wormhole (see \cite{Olmo:2015bya} for more details).  Using (\ref{eq:geoBInear}) and (\ref{eq:Vzero}), one finds that the affine parameter behaves as 
\begin{equation}\label{eq:affineNull}
\lambda(x)\approx  \lambda_0\pm \frac{x}{3}\left|\frac{x}{a}\right|^{\frac{1}{2}}\left(1 -\frac{3(b+E^2)}{10}\left|\frac{x}{a}\right| \right)\ .
\end{equation}
This solution (which is unique) guarantees the extendibility of the affine parameter accross $x=0$. Therefore, all time-like and null geodesics in these space-times are complete regardless of the existence of curvature divergences at the wormhole throat.

\subsection{Geodesics in $f(R)$ gravity}

In the $f(\Rpal)$ case, our general approach for the description of geodesics leads to the following equation
\begin{equation}\label{eq:geofR}
\frac{1}{f_{\Rpal}^2}\left(\frac{dx}{d\lambda}\right)^2=E^2-\frac{A(x)}{f_{\Rpal}}\left(\frac{L^2}{r^2(x)}-k\right) \ .
\end{equation} 
Let us consider first the case with $0<\gamma< 1$, for which there is no wormhole structure. In these cases,  as $x\to 0$ we find $f_{\Rpal}\approx (1-\gamma)$, $r(x)\approx x/\sqrt{1-\gamma}$, and
\begin{equation}
A(z)\approx1-\frac{r_S (\delta_c^{(\gamma)}-\delta_1)}{r_c \delta_c^{(\gamma)}\sqrt{1-\gamma} \ z}-\frac{r_S\delta_1}{2r_c}z^2+\ldots 
\end{equation}
With this, near the center (\ref{eq:geofR}) can be written as
\begin{equation}\label{eq:geofR1}
\left(\frac{dr}{d\lambda}\right)^2=\tilde{E}^2-A(r)\left(\frac{L^2}{r^2}-k\right) \ ,
\end{equation} 
with $\tilde{E}^2=(1-\gamma)E^2$. The discussion now proceeds in much the same way as in models of non-linear electrodynamics coupled to GR. One can find configurations for which the metric is regular at the origin, $\delta_c^{(\gamma)}=\delta_1$, and others with divergences, $\delta_c^{(\gamma)}\neq\delta_1$. A detailed discussion of geodesics in such  configurations  will be provided elsewhere \cite{Bejarano}. The key point to note here is that the absence of a wormhole implies that  radial null geodesics, $\left(\frac{dr}{d\lambda}\right)^2=\tilde{E}^2$, always reach $r=0$ in a finite proper time with no possibility of extension beyond that point. Thus, similarly as in the Reissner-Nordstr\"{o}m case of GR, such solutions can be regarded as singular. \\

Let us now consider the case with $\gamma>1$, for which there is a wormhole. From previous results, we know that as the wormhole is approached, we have $A(x)\approx \tilde{C}/(z-z_c)$ and $f_{\Rpal}\approx  \frac{8z_c^3}{1+z_c^4}(z-z_c)$, which implies that the right-hand side of (\ref{eq:geofR}) must vanish at some $z>z_c$ if $L\neq 0$ or $k=-1$ (time-like observers). This means that such geodesics never reach the wormhole throat, which is similar to what we already observed in the case of Reissner-Nordstr\"{o}m in GR, where time-like observers and $L\neq0$ geodesics never reach the center. If we consider radial null geodesics, (\ref{eq:geofR}) turns into
\begin{equation}\label{eq:geofR2}
\frac{1}{f_{\Rpal}^2}\left(\frac{dx}{d\lambda}\right)^2=E^2\ .
\end{equation} 
Far from the wormhole $f_{\Rpal}\to 1$ and this recovers the standard behavior $r\approx x\approx \pm E(\lambda-\lambda_0)$, with the $+/-$ sign corresponding to outgoing/ingoing rays. Now, near the wormhole, we can use the relation $r^2 f_{\Rpal}=x^2$ and the fact that $r\to r_c$ as $x\to 0$ to write (\ref{eq:geofR2}) as
\begin{equation}\label{eq:geofR3}
\frac{r_c^4}{x^4}\left(\frac{dx}{d\lambda}\right)^2=E^2\ ,
\end{equation} 
which leads to 
\begin{equation}
-\frac{1}{x}=\pm \frac{E}{r_c^2}(\lambda-\lambda_0) \ .
\end{equation}
From this it follows that as $x\to 0$, $\lambda\to -\infty$ for outgoing rays, while for ingoing rays $\lambda\to +\infty$. Stated in words, ingoing rays which started their trip from $x\to +\infty$ and $\lambda\to -\infty$ approach the wormhole at $x\to 0$ as $\lambda\to +\infty$, whereas outgoing rays which started their trip near the wormhole at $\lambda\to -\infty$ propagate to infinity as  $\lambda\to +\infty$. Thus, all time-like and null geodesics in these configurations ($\gamma>1$) are complete. Curvature divergences, which arise at the wormhole throat, cannot be reached in a finite affine parameter and, therefore, do not belong to the physically accessible region. These solutions are nonsingular even though one can never go through the wormhole. If one considers the region $x<0$, identical conclusions are obtained. 

\section{Summary and conclusions}\label{sec:theend}

In these Lectures we have studied the classical problem of black hole singularities from a four dimensional geometric perspective. Motivated by the fact that GR predicts the existence of singularities in simple static, spherically symmetric configurations, we have considered extensions of the theory to test the robustness of this disturbing result. In our study we have not followed the traditional  approach of implicitly assuming that the space-time geometry is Riemannian. Rather, we have emphasized that the type of geometry associated with the gravitational interaction is an empirical question that must be settled by experiments, not imposed by convention or tradition. Whether the geometry is Riemannian or not is as fundamental a question as the number of space-time dimensions or the existence of supersymmetry, which are aspects that have received much attention in the last years. 

We have thus considered a metric-affine geometrical framework for the formulation of our extensions of GR, with the additional simplification of setting torsion to zero (Palatini approach \cite{Olmo:2011uz,Origin}). This choice is justified on simplicity grounds, as a first step in the exploration of new gravitational physics. The inclusion of fermionic matter, whose spin sources the torsion, would require a detailed treatment beyond the Palatini approach. \\

An unusual property of the gravity theories considered here, as compared to the more standard metric or Riemannian approach, is that their modified dynamics arises as a result of nonlinearities generated by the matter fields rather than by the emergence of new dynamical degrees of freedom. In fact, the field equations of $f(\Rpal)$ theories, the Born-Infeld model, or any Lagrangian which is just a function of the inverse metric and the Ricci tensor \`{a} la Palatini admit a generic representation that exactly recovers the equations of GR (with an effective cosmological constant) in vacuum when the matter fields are absent \cite{Olmo:2012yv,Jimenez:2015caa,Bazeia:2015zpa,ERE2011}. This means that generically these theories neither exhibit ghosts nor massive gravitons. These properties together with the second-order character of the field equations should be regarded as general characteristics of the metric-affine formulation. \\

In our opinion, the most remarkable aspect of the theories presented here is that they do what they were expected to do in a simple and clean manner. They were conceived as extensions of GR which could bring new relevant physics at high energies, and they yield solutions which are in agreement with GR almost everywhere, except in regions of very high energy density. The modifications that they introduce are such that black hole centers acquire a nontrivial structure that allows to preserve the completeness of geodesics. In the Born-Infeld type model, geodesics can go through the central wormhole, whereas in the $f(\Rpal)$ case, the wormhole (when it exists) lies beyond the reach of the geodesics. \\

Following the standard definition of space-time singularities given in the specialized literature and main text books on gravitation \cite{Geroch:1968ut,Hawking:1973uf,Wald:1984rg,Senovilla:2014gza}, we have concluded that the solutions containing wormholes are nonsingular because they are geodesically complete. And this is so despite the appearance of curvature divergences at the wormhole throat. One should note, however, that there exists a widespread tendency in the literature to simplify the complex notion of space-time singularity and associate the divergence of certain quantities (such as curvature scalars or tensor components) with its definition. This tendency can be partly justified by the {\it strong correlation} existing between the occurrence of divergences and the incompleteness of some geodesics. Somehow, one tends intuitively to associate divergences with geodesic incompleteness as if the former were the cause/reason for the latter \cite{Curiel2009}. We have shown here with several explicit examples that black hole space-times can be geodesically complete and at the same time have curvature divergences, thus breaking the correlation typically found in GR. \\

Divergences in curvature tensors/scalars are obviously associated with strong tidal forces. The effects of such forces have been investigated in the literature by means of geodesic congruences in an attempt to classify the strength of singularities \cite{Ellis,Tipler,CK,Nolan,Ori,Tipler77,Nolan:2000rn}. In that context, extended physical objects are represented as congruences of geodesics, and the evolution of their relative distance as curvature divergences are approached provides information about their fate. Those methods have been applied in  the general charged solutions of the Born-Infeld model studied here finding that the different parts of a body that goes through the wormhole never lose causal contact among them despite the existence of infinite accelerations at the throat \cite{AGDip}. This offers a new view on the problem which should be further investigated to better understand if curvature divergences possess any {\it destructive} power. We would like to emphasize that though in the Born-Infeld model physical observers do interact with the curvature divergence as the wormhole is crossed, in the $f(\Rpal)$ case, the divergence is never reached in a finite affine distance. Therefore, the $f(\Rpal)$ model is free from the potential drawbacks of directly interacting with a curvature divergence, as it lies beyond the physically accessible space-time.\\

Though much research is still needed to better understand gravitational and non-gravitational physics in metric-affine spaces, the point is that two analytically tractable {\it toy models} with nontrivial results about black holes are already available. \\

Before concluding, we must note that our approach has assumed that particles and observers can be viewed as structureless entities (geodesics), whereas physical measurements are carried out by means of probes with wave-like properties because matter fields are of a quantum nature. One should thus study the propagation of waves in these space-times to see how they behave and interact with regions of intense gravitational fields such as wormhole throats, where curvature scalars typically diverge. A first analysis in this direction was carried out in \cite{Olmo:2015dba}, where the scattering of scalar waves in horizonless (naked) configurations was considered. Despite the infinite potential barrier that curvature divergences generate, one verifies that the propagation through the wormhole is smooth and that transmission and reflection coefficients can be computed numerically and contrasted with analytical estimates, yielding good agreement. These results, therefore, give further support to the absence of singularities in these geometries. \\

\begin{acknowledgement}
The author is supported by a Ramon y Cajal contract, the Spanish grants FIS2014-57387-C3-1-P and FIS2011-29813-C02-02 from MINECO, the grants i-LINK0780 and i-COOPB20105 of the Spanish Research Council (CSIC), the Consolider Program CPANPHY-1205388, and the CNPq project No. 301137/2014-5 (Brazilian agency). 
\end{acknowledgement}

\end{document}